\documentclass[namedreferences]{solarphysics}
\usepackage[hyperref,optionalrh]{spr-sola-addons} 
\usepackage{epsfig}          
\usepackage{graphicx}        
\usepackage{color}           
\usepackage{breakurl}        

\newcommand{\kms}{{\rm km\,s$^{-1}$}}

\newcommand{\rsun}{{R$_\odot$}}
\newcommand{\arcsec}{{$^{''}$}}
\newcommand{\degree}{{$^{\circ}$}}

\chardef\us=`\_

\begin{document}
\begin{article}
\begin{opening}

\title{Population of Bright Plume Threads in Solar Polar Coronal Holes}

\author[addressref={aff1},corref,email={z.huang@sdu.edu.cn}]{\inits{Z.}\fnm{Zhenghua}~\lnm{Huang}\orcid{0000-0002-2358-5377}}
\author[addressref=aff2]{\inits{Q.}\fnm{Quanhao}~\lnm{Zhang}\orcid{0000-0003-0565-3206}}
\author[addressref=aff1]{\inits{L.}\fnm{Lidong}~\lnm{Xia}\orcid{0000-0001-8938-1038}}
\author[addressref=aff3]{\inits{L.}\fnm{Li}~\lnm{Feng}}
\author[addressref=aff1]{\inits{H.}\fnm{Hui}~\lnm{Fu}\orcid{0000-0002-8827-9311}}
\author[addressref=aff1]{\inits{W.}\fnm{Weixin}~\lnm{Liu}}
\author[addressref=aff1]{\inits{M.}\fnm{Mingzhe}~\lnm{Sun}}
\author[addressref=aff1]{\inits{Y.}\fnm{Youqian}~\lnm{Qi}}
\author[addressref=aff1]{\inits{D.}\fnm{Dayang}~\lnm{Liu}}
\author[addressref=aff3]{\inits{Q.}\fnm{Qingmin}~\lnm{Zhang}\orcid{0000-0003-4078-2265}}
\author[addressref=aff1]{\inits{B.}\fnm{Bo}~\lnm{Li}\orcid{0000-0003-4790-6718}}
\address[id=aff1]{Shandong Key Laboratory of Optical Astronomy and Solar-Terrestrial Environment, Institute of Space Sciences, Shandong University, Weihai 264209, Shandong, China}
\address[id=aff2]{CAS Key Laboratory of Geospace Environment, School of Earth and Space Sciences, University of Science and Technology of China, Hefei 230026, Anhui, China}
\address[id=aff3]{Key Laboratory for Dark Matter and Space Astronomy, Purple Mountain Observatory, CAS, Nanjing 210034, China}
\runningauthor{Z. Huang et al.}
\runningtitle{Population of Bright Plume Threads}

\begin{abstract}
Coronal holes are well accepted to be source regions of the fast solar wind.
As one of the common structures in coronal holes, coronal plumes might contribute to the origin of the nascent solar wind.
To estimate the contribution of coronal plumes to the nascent solar wind, we make the first attempt to estimate their populations in the solar polar coronal holes.
By comparing the observations viewed from two different angles taken by the twin satellites of STEREO and the results of Monte Carlo simulations, we estimate about 16--27 plumes rooted in an area of $4\times10^4$\,arcsec$^2$ of the polar coronal holes near the solar minimum, which occupy about 2--3.4\% of the area.
Based on these values, the contribution of coronal plumes to the nascent solar wind has also been discussed.
A further investigation indicates that more precise number of coronal plumes can be worked out with observations from three or more viewing angles.
\end{abstract}


\end{opening}

\section{Introduction}
\label{sect_intr}
Coronal holes (CHs) are dark areas in the solar corona as viewed in EUV and soft X-ray passband.
They are (almost) permanent structure in the polar region, but can also be seen in any other location on the Sun\,\citep{2009LRSP....6....3C}.
It is believed that CHs are sources of fast solar wind, the high speed ($>$500\,\kms) component of the stream of plasma in the interplanetary space flowing outward from the Sun\,\citep{1973SoPh...29..505K}.
While slow solar wind might come from active regions\,\citep[see e.g.][]{2020ApJ...894..144B},
brightenings occurring at the boundaries of CHs also play an important role\,\citep[][]{2009A&A...503..991M,2010A&A...516A..50S,2012ApJ...750...50N,2012A&A...545A..67M}.
Thus, CHs have been intensively studied in order to understand the formation of the nascent solar wind\,\citep[e.g.][]{1999Sci...283..810H,2003A&A...399L...5X,2004A&A...421..339P,2004A&A...424.1025X,2005Sci...308..519T,2010AdSpR..45..303H}.

\par
\begin{figure*}
\centering
\includegraphics[clip,trim=1cm 0.5cm 2cm 1.2cm,width=\linewidth]{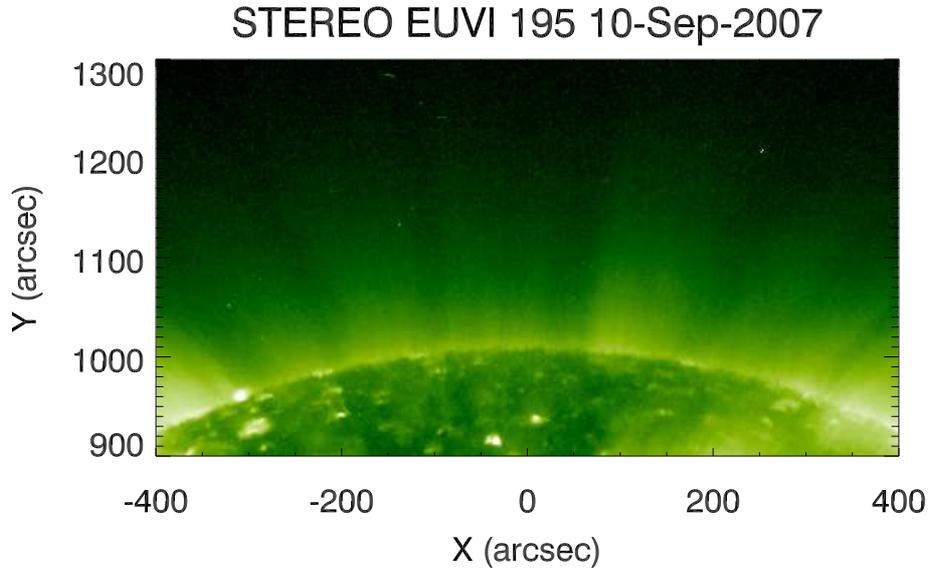}
\caption{
The north polar cap of the Sun seen in EUV 195\,\AA\ passband taken by SECCHI/EUVI on-board the STEREO-A satellite.
Coronal plumes are ray-like bright features extending outward the solar limb.
In this image, one arc-second (arcsec) is equivalent to 696\,km on the Sun.
}
\label{fig:example}
\end{figure*}

Although CHs are relatively ``quiet'' compared to the other places of the solar corona, they are structured with bright points\,\citep[e.g.][]{2012A&A...548A..62H,2016ApJ...818....9M}, coronal jets\,\citep[e.g.][]{2011ApJ...735L..43S,2018ApJ...853..189P,2019ApJ...873...93K,2019ApJ...882...16M} and coronal plumes\,\citep[e.g.][]{1977SoPh...53..397A,2011SCPMA..54.1906Y,2014ApJ...787..118R}, etc.
Coronal plumes (CPs) are bright ray-like features that are abundant in CHs\,\citep{2011A&ARv..19...35W,2015LRSP...12....7P}.
CPs can be easily seen in the polar CHs where they extend off-limb and thus have a great contrast to the background (see Figure\,\ref{fig:example}).
Many studies\citep[e.g.][]{1998ApJ...501L.145W,2008SoPh..249...17W,2009ApJ...700..551G,2014ApJ...793...86P} have found that CPs in CHs are normally rooted in coronal bright points that are small compact brightenings in the corona with a diameter of 20--30 Mm in average\,\citep{2019LRSP...16....2M} or network regions that are lane-like structures in the ``quiet'' solar chromosphere consisting of numerous fine-scaled active bright dots.

\par
Because CPs usually occur in CHs and apparently extend along the open magnetic field lines, their connection with solar wind has been a major interest of the community. 
By studying spectroscopic data with the Doppler dimming technique, \citet{2003ApJ...589..623G} found that CPs have outward velocities in excess of 60\,\kms\ at the heights ranging from 1.05 to 1.35\,\rsun (solar radii) and they estimate that CPs contribute about half of the mass of the fast solar wind at 1.1\,\rsun.
A further study\,\citep{2005ApJ...635L.185G} suggests that the mass in plumes might transfer to inter-plume regions while they are propagating outward exceeding 1.6\,\rsun.
More recently, analyses of on-disk plumes revealed that their Doppler velocities increase with heights from 10\,\kms\ at 1.02\,\rsun\ to 25\,\kms\ at 1.05\,\rsun, suggesting a significant total of mass balancing the loss by the solar wind\,\citep{2014ApJ...794..109F}.
Using spectral data from UVCS/SOHO and imaging data from LASCO/SOHO, \citet{2020A&A...643A.104Z} conclude that CPs can contribute about 20\% of the solar wind originated from the solar poles.
Furthermore, propagating quai-periodic disturbances whose speeds are more than 100\,\kms\ usually present in CPs, and they might also play an important role in carrying mass and/or energy from lower solar atmosphere to corona and then power the solar wind\,\citep{1998ApJ...501L.217D,2000ApJ...533.1071O,2010ApJ...718...11G,2011ApJ...736..130T,2011ApJ...738...18T,2011A&A...528L...4K,2012ApJ...759..144T,2015ApJ...809L..17J,2019SoPh..294...92Q,2020ApJ...900L..19C}.
However, the contributions of CPs to solar wind is still under debate.
Some studies found that inter-plume regions contribute the major part of the mass flowing in the fast solar wind from coronal holes\,\citep{2000A&A...353..749W,2000A&A...359L...1P,2000ApJ...531L..79G,2003ApJ...588..566T,2009ApJ...700..292F}.

\par
In order to assess the contribution of CPs to the solar wind and/or coronal heating, it is crucial to estimate the ratio of the area occupied by CPs to the whole CH (i.e. the population of CPs).
This is especially difficult in the polar region because CPs' emission is optically-thin and the line-of-sight effect brings any CPs in the background and foreground to the same imaging plane as exampled in Figure\,\ref{fig:example}.
As concluded by\,\citet{2003ApJ...589..623G},  CPs apparently make a substantial contribution to the line-of-sight intensity.
Such estimation to the population of CPs is impossible-to-some-extent using observations from a single viewing-angle,
but a solution might be given by multi-angle observations, which have been successfully used to determine the 3D geometry of plumes\,\citep{2009ApJ...700..292F,2013SoPh..283..207D} and coronal loops\,\citep{2007ApJ...671L.205F,2007SoPh..241..235F,2009ApJ...695...12A,2011LRSP....8....5A,2013ApJ...763..115A}, 3D locations of comets\,\citep{2020ApJ...897...87C} and 3D structures of solar winds\,\citep{2018JGRA..123.7257L,2020AdSpR..66.2251L,2020JGRA..12527513L} and coronal mass ejections\,\citep{2012ApJ...751...18F,2013SoPh..282..221F}, etc.
Here, using observations from the twin STEREO (Solar TErrestrial RElations Observatory) satellites\,\citep[STEREO-A and STEREO-B,][]{2008SSRv..136....5K}, we make the first attempt to estimate the population of CPs in the solar polar coronal holes.

\par
In what follows, we describe observations in Section\,\ref{sect_method} and methodology and results in Section\,\ref{sect_res}, give prospects in Section\,\ref{sect_pro} and conclusion in Section\,\ref{sect_con}.

\begin{table}
\caption{Details of the data analysed in the present study and the estimated plume populations based on these observations.}\label{tab_data}
\begin{tabular}{ccccccc}
  Date & Time &Separation\tabnote{Separation angle between STEREO-A and STEREO-B in HEEQ coordinates.} & D$_A$ /D$_B$\tabnote{D$_{A(B)}$ is the distance between STEREO-A(B) and the Sun.} & Number of&Occupancy \\
  YYYYMMDD&(UT) &  (\degree) & & CP threads\tabnote{Number of CP threads estimated in an area of 200\arcsec$\times$200\arcsec\ above polar coronal holes.}&of CPs\tabnote{Occupancy in area based on the number of CP threads. See the main text for detail.}\\
  \hline
20070910&00:05&30.1&0.888&22&2.8\%\\
20080710&00:05&60.2&0.891&27&3.4\%\\
20090201&00:05&90.4&0.957&20$\sim$30&2.5\%$\sim$3.8\%\\
20091012&00:05&120.2&0.890&16&2.0\%\\
20100805&00:05&150.2&0.905&20&2.5\%\\
\hline
\end{tabular}
\end{table}

\section{Observations}
\label{sect_method}
The observations analysed here were achieved by the Extreme Ultraviolet Imager\,\citep[EUVI,][]{2004SPIE.5171..111W} that is part of the Sun Earth Connection Coronal and Heliospheric Investigation\,\citep[SECCHI,][]{2008SSRv..136...67H} on-board STEREO.
They are subregions of the images with a circular field-of-view upto $\pm$1.7\, solar radii taken at the 195\,\AA\ passband\,\citep[representative of temperatures peak at 1.6\,MK,][]{2008SSRv..136...67H}.
The angular sampling of the images is 1.6\arcsec\ per pixel.
The data include those taken from both the twin satellites of STEREO-A and STEREO-B,
on which the EUVI telescopes are designed to be identical.
We used the data when the twin satellites are separated for roughly 30\degree, 60\degree, 90\degree, 120\degree\ and 150\degree\ in the Heliocentric Earth EQuatorial coordinate system\,\citep[HEEQ,][]{2006A&A...449..791T}.
The details of the analysed data are given in Table\,\ref{tab_data}.

\par
The data are calibrated with the standard procedure of \textit{secchi\_prep.pro} provided in the \textit{solar software} by the instrument team.
We then use the procedure of \textit{scc\_roll\_image.pro} to roll and align the images to the Radial-Tangential-Normal (RTN) coordinate system that put the solar north up\,\citep{2006A&A...449..791T,2010SoPh..261..215T}.
In order to directly compare the two images observed by the two satellites at a given time, the ratio of the distances between the Sun and the satellite for STEREO-A and STEREO-B (see Table\,\ref{tab_data}) are used to reconstruct the images for a pixel of the two images having the same spatial scale on the Sun.

\section{Methodology and Results}
\label{sect_res}
As mentioned earlier, the line-of-sight effect is the key problem that brings difficulty in determining the population of CPs in the polar CH.
More than one distributions of CPs can lead to the same observables at a single viewing angle.
Thus simultaneous observations from an additional viewing angle can provide extra constraints and potentially give a solution.
To find such a solution for achieving the population of CPs, we develop a method based on Monte Carlo simulations.
The simulations randomly provide distributions of CPs in a selected region of the solar polar CH, and compute the observables at the corresponding viewing angles.
The randomness of the distribution includes the number of CPs and the locations where the CPs rooted in a CH.
The correlation coefficients between the simulated observables and those observed by the two satellites are then calculated.
The distributions of CPs having correlation coefficients better than 2$\sigma$ (hereafter, correlation coefficient threshold) above the average of all cases for both viewing angles are then selected as potential solutions.

\par
\begin{figure*}[t]
\centering
\includegraphics[clip,trim=1.5cm 0.5cm 2.4cm 0.5cm,width=\linewidth]{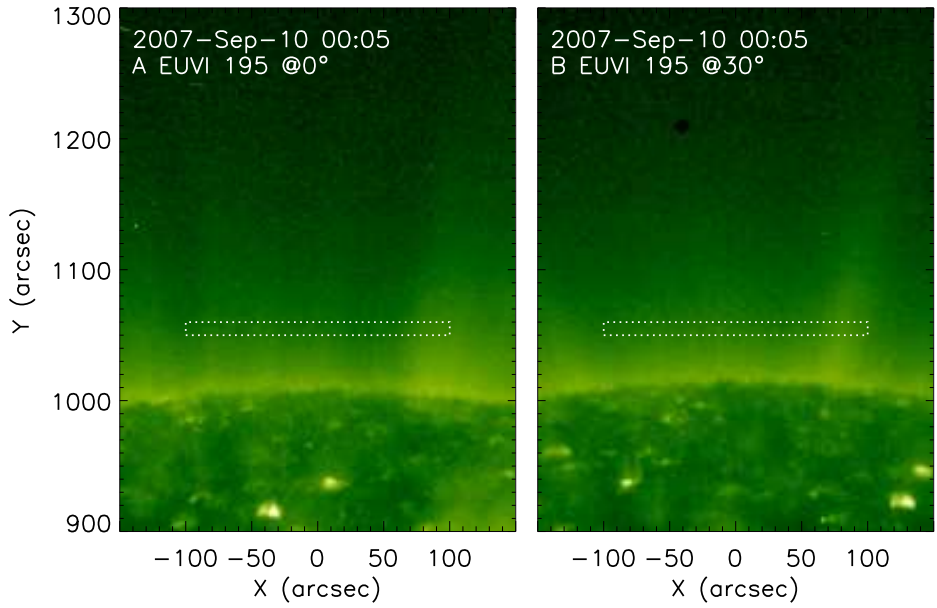}
\caption{
The north polar cap of the Sun in EUV 195\,\AA\ passband observed by SECCHI/EUVI on-board the STEREO-A (left) and STEREO-B (right) while the two satellites are separated by about 30\degree.
A radial filter has been applied on the regions above the limb and the images are shown in logarithm scale.
This is an example of the polar region of the Sun studied in the present study.
The region enclosed by the dotted lines is used to determine the fine structures of the polar plumes.
}
\label{fig:fov}
\end{figure*}

As the first step, we produce the intensity variation curve (IVC) along solar\_X from --100\arcsec\ to +100\arcsec\ (the unit of arcsec is based on STEREO-A) averaging in the solar\_Y range of 50--60\arcsec\ above the north solar limb (see the dotted lines in Figure\,\ref{fig:fov}).
Such IVCs are obtained from observations taken from both A and B satellites.
Because the studied region is off-limb and centering at solar\_X=0, 
the IVCs from both satellites are supposed being from the field-of-views having the same central point.
The IVCs are then smoothed with a window of 3 to increase signal-to-noise ratio (see black curves as example in Figure\,\ref{fig:ivcs}).
Next, we identify the local peaks (humps) in the IVC using the method on the basis of derivative of the curves as described in previous studies\,\citep{2017MNRAS.464.1753H}.
The selected solar\_Y locations for obtaining the IVC  is far from the limb where the intensity of EUVI 195\,\AA\ drops
 to about $1/e$ of the maximum of that at the limb.
We have also applied the same procedures on the location range of 90--100\arcsec above the limb, and we obtain similar values (results not shown).
We believe that the identified local peaks are corresponding to bright plume threads 
since most of the omnipresent spicules can only reach a height of about 20\arcsec\,\citep[see][and references therein]{2000SoPh..196...79S}.
An array of factors is then determined as the round values of the intensities of the local peaks divided by the minimum of all.
For each IVC, we generate a clean curve having the same elements as the original one, in which the locations of local peaks are given the obtained factors and the rest are given as zero (see the orange lines in Figure\,\ref{fig:ivcs}, where 19 peaks are identified in the STEREO-A curve and 20 in the STEREO-B curve).
Here we denote the numbers of local peaks identified from the STEREO-A and STEREO-B observations as $N_A$ and $N_B$, respectively.
We then run Monte Carlo simulations by letting the number of plume threads vary from the maximum of [$N_A$,$N_B$] to $N_A\times N_B$ and letting the corresponding plumes randomly locate in an area of solar\_X=[--100,100] and solar\_Z=[-100,100] (solar\_Z is along line-of-sight at STEREO-A viewing angle).
For each selected number of plumes, the simulations consider 10,000 random combinations of their locations. 
In total, a Monte Carlo simulation for a specific separation of STEREO-A and STEREO-B includes $10^4\times(N_A\times N_B - max([N_A,N_B]) +1)$\ random cases, which is $3.61\times10^6$ for the example shown in Figure\,\ref{fig:ivcs}.
In the simulations, each plume is assumed to have intensity of 1.
If there are more than one plumes in the same line-of-sight, the observables sum all of them based on the optically-thin assumption.
We can then obtain the simulated IVCs based on a specific viewing angle as determined by the separation of STEREO-A and STEREO-B.
The correlation coefficients of the simulated IVCs and the corresponding clean curves based on observations are calculated to determine which plume population(s) can best reproduce the observations.
If correlation coefficients based on both STEREO-A and STEREO-B observations meet the criterion mentioned earlier, we denote the corresponding guess as a hit.

\par
\begin{figure*}
\centering
\includegraphics[clip,trim=0.5cm 1cm 0.5cm 1cm,width=\linewidth]{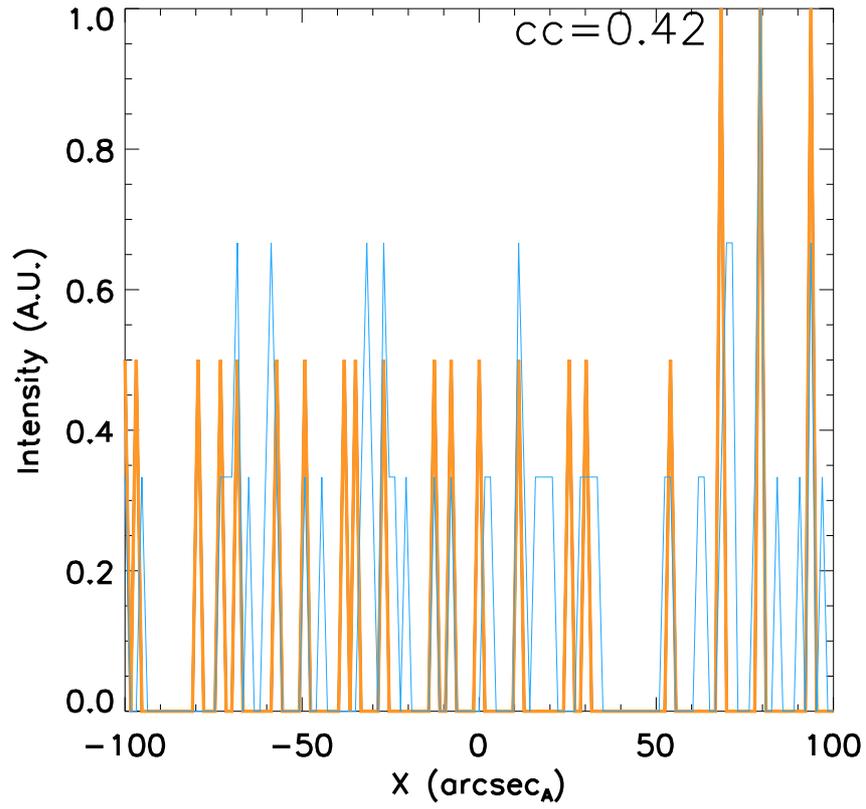}
\caption{
A comparison to a simulated curve (blue line) and the observed curve (orange line). The correlation coefficient between the two curves is 0.42.
}
\label{fig:lcs_exmp}
\end{figure*}

In Figure\,\ref{fig:cchist030}, we give histograms of the number of hits varying with the guessed number of plumes.
We also show in this case the results while consider 100,000 random distributions of locations of plumes for the simulation (see the orange curves in Figure\,\ref{fig:cchist030}).
We can see that the histograms from 10,000 and 100,000 random distributions of locations are giving almost identical results in relative.
Therefore, here we simulate only 10,000 cases in order to reduce the consumption of the computing resources.
From Figure\,\ref{fig:cchist030}, we can see that a large range of numbers of plumes can reproduce the observations, which is consistent with one's common sense.
It is also shown in the histogram that the distribution peaks at a certain guessed number of plumes.
We take this guess of number of plumes as the most possible number of plumes in the reality.
For the example shown in Figure\,\ref{fig:cchist030}, that is 22.
As one can see that this number is close to the number of humps identified in the IVCs, it indicates that CPs are sparse in the polar CH.
We would like to note that the correlation coefficient computed here is linear Pearson correlation.
Because the curves used for calculations are ``clean'' with only spikes, such correlation is very sensitive if there is any shift in those spikes, and thus most of the results obtained here are less than 0.5.
Even though such correlation coefficients are small, the data numbers are large (125 points in these cases) and thus it is still significant that the two data series are correlated (also see an example in Figure\,\ref{fig:lcs_exmp}).
Actually, we have tested the method on artificial data before applied on the observations (see an example shown in the appendix).
We found that it is acceptable based on two viewing angles in cases where plumes are not too dense.

\par
In Figure\,\ref{fig:cchist}, we show the histograms of the number of hits based on the observations while the two satellites are separated by 60\degree, 90\degree, 120\degree and 150\degree.
The best estimated numbers of plumes for the observations at separations of 60\degree,120\degree and 150\degree\ are 27, 16 and 20, respectively.
Although the simulations based on the observations taken at 90\degree\ give the best estimation at the plume number of 61, we see that the numbers of hits are generally higher in the range between 20 and 30, and thus it is reasonable to assume the correct number falling in this range too.
These values indicate that plume populations based on these five datasets are quite close, and this is reasonable as all the observations were taken at the time near the solar minimum.

\par
In Figure\,\ref{fig:ivcs}, we found that some plume threads seemingly locate next to each other on the plane of sky, so that their corresponding peaks in the lightcurves are very close to each other (see the region denoted by the dotted line in Figure\,\ref{fig:ivcs}.
It is reasonable to assume that the total of the cross-sections of these plume threads is equivalent to the spatial range in X between the two dips at both sides (see the dotted lines in Figure\,\ref{fig:ivcs}).
Based on this assumption, we obtain a good estimation of about 8\arcsec\ for the cross-sections of the plume threads,
which is consistent with that reported in the previous study\,\citep{1997SoPh..175..393D}.
On the assumption that these plume threads have a cylinder geometry,
we can obtain the total area of the solar surface that all the plume threads occupy.
Given 20 plume threads, the area is estimated to be about 1,000\,arcsec$^2$.
Compared to the total area of CH of the observation and simulations, which is 40,000\,arcsec$^2$, plumes appear to occupy only a small fraction ($\sim$2.5\%) of the area in the CH.
The occupancies of CP threads based on the four data sets range from 2\% to 3.4\% (see Table\,\ref{tab_data}).
Please note that these results can be affected by a few factors, including the resolution and sensitivity of the instruments and the relevant assumptions in our method.
For example, the method does not take into account the variety of the width and emission of the plume threads.
Therefore, these numbers can be treated as the lower limits of the estimations.
Since CPs are rooted in the network lanes, this proportion is half of that of the size of network magnetic elements (representative of the network lane) to a supergranulation (representative of the full network cell), in which the network magnetic elements have a size in the order of 1\arcsec\,\citep{2007ApJ...666..576D,2008ApJ...674..520L,2009A&A...505..801X,2010ApJ...720.1405L,2020ApJS..250....5J} and the supergranulations have a size of about 30--40\arcsec\,\citep{2010LRSP....7....2R} and both can be assumed to be circular geometries.
This suggests that not all network lanes produce CPs, which is consistent with observations\,\citep{2019SoPh..294...92Q}. 
By taking into account the density of CPs is typically about five times of that of the inter-plume regions\,\citep{2011A&ARv..19...35W}, CPs can contribute about  9--15\% of the solar winds from the solar poles, which is less than the recent result reported by \citet{2020A&A...643A.104Z}.
The obtained values were not taken into account the uncertainty brought in by the method and the difference between the speeds of plasma flows in CPs and that in the inter-plume regions.
In the near future, spectroscopic observations of polar CHs from out of the ecliptic plane by Solar Orbiter\,\citep{2020A&A...642A...1M,2020A&A...642A..14S} shall provide direct measurement of the outflow structures in the polar CHs and thus will give a more precise estimation to the contribution from CPs to the nascent solar wind.

\par
Moreover, plumes are also abundant in on-disk coronal holes\,\citep[e.g.][]{2016ApJ...818..203W,2018ApJ...861..111A}.
The estimation of area of plumes in coronal holes can be improved by using observations of on-disk coronal holes.
This will be considered in a future study by taking into account additional treatments such as background subtraction and projection effect.

\section{Prospects}
\label{sect_pro}
In the present work, we make the first attempt to estimate the population of coronal plume threads in the polar coronal hole based on Monte Carlo simulations and observations taken by the twin STEREO satellites at two different viewing angles.
As we can see from the results, the solutions given by the simulation have to be determined from the peaks of the histograms of hits.
A question is whether an extra constraint (i.e. observations from the third viewing angle) can help giving a better solution. 
All these constraints should be taken with identical instruments like STEREO-A\&B, otherwise more free variables should be taken into account.
In the appendix, we have tested a case with sparse plumes and found that observations from three viewing angles give much more promising results.
Recently, \citet{2020ScChE..63.1699W,wang2020scic} propose a concept mission, namely ``Solar Ring'', for observing the Sun simultaneously from six different viewing angles with duplicated instruments, and we believe such observations shall be powerful in solving the issue here.

\par
\begin{figure*}
\centering
\includegraphics[clip,trim=1cm 1cm .3cm .3cm,width=\textwidth]{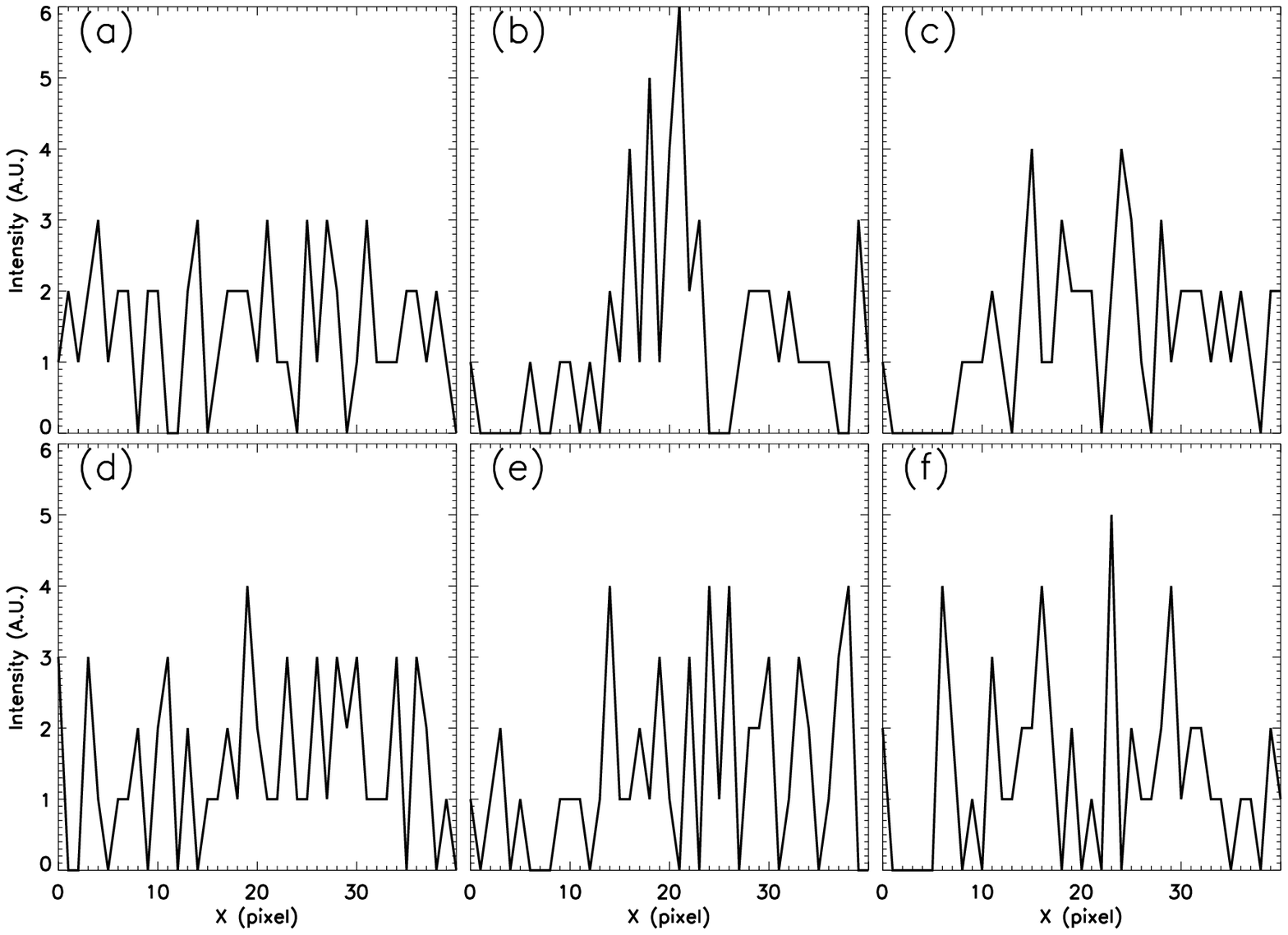}
\caption{The intensity variation curves of the artificial data with 60 plumes in a region of 40\,pixels$\times$ 40\,pixels viewed at 0\degree\ (a), 30\degree\ (b), 60\degree\ (c), 90\degree\ (d), 120\degree\ (e) and 150\degree\ (f).
}
\label{fig:lcs_sim}
\end{figure*}
\begin{figure*}
\centering
\includegraphics[clip,trim=0cm 0cm 0cm 0cm,width=0.48\textwidth]{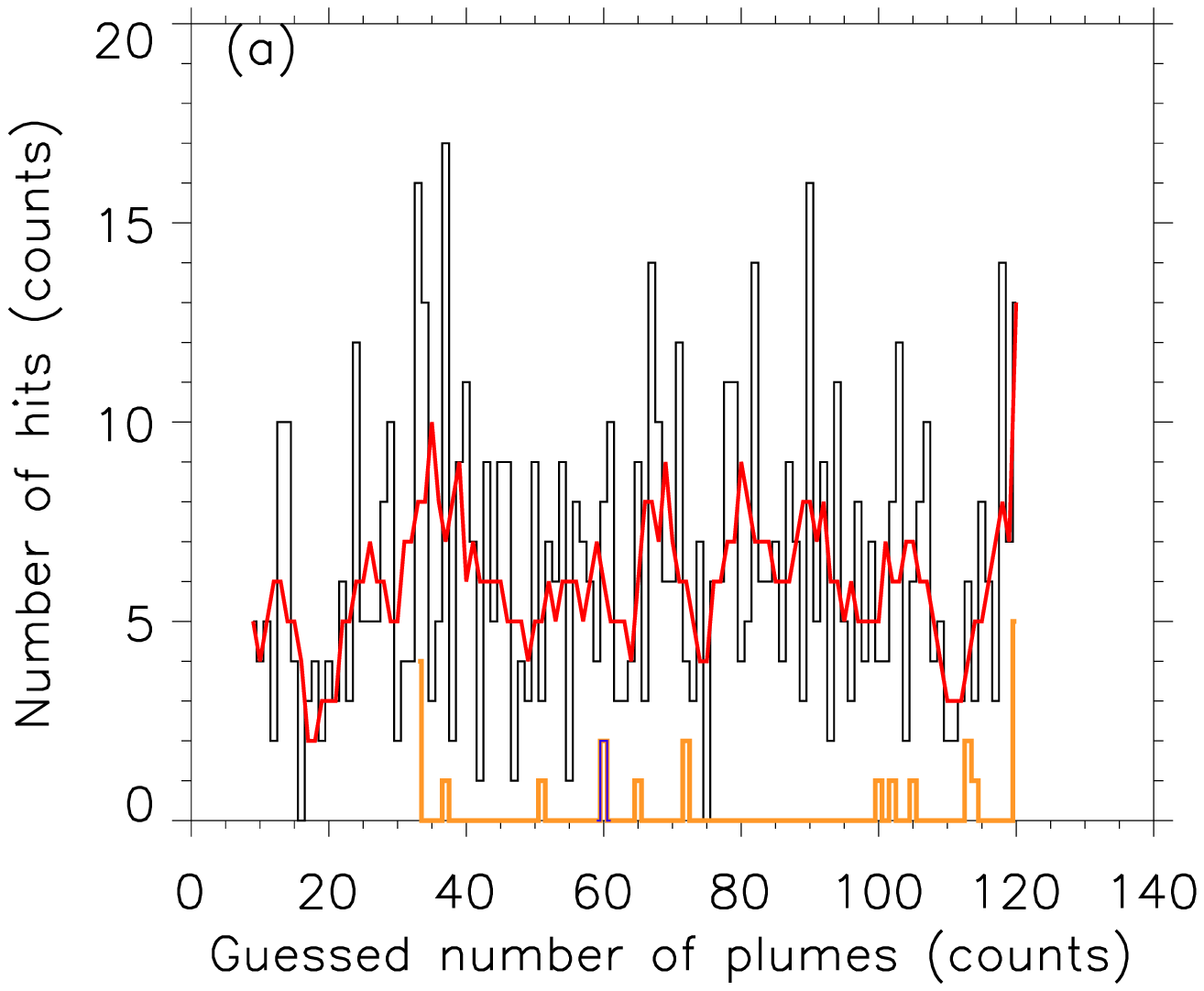}
\includegraphics[clip,trim=0cm 0cm 0cm 0cm,width=0.48\textwidth]{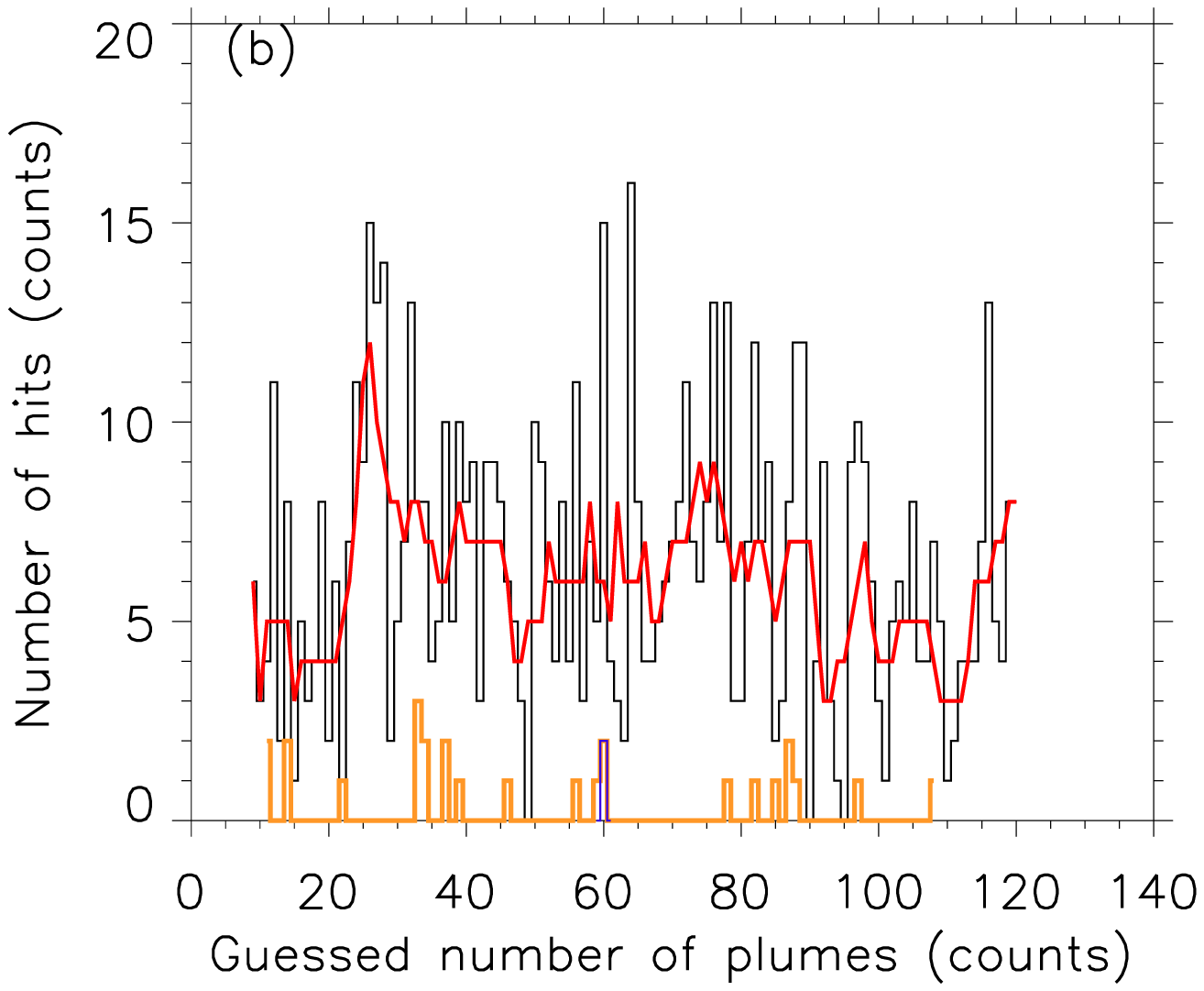}\\
\includegraphics[clip,trim=0cm 0cm 0cm 0cm,width=0.48\textwidth]{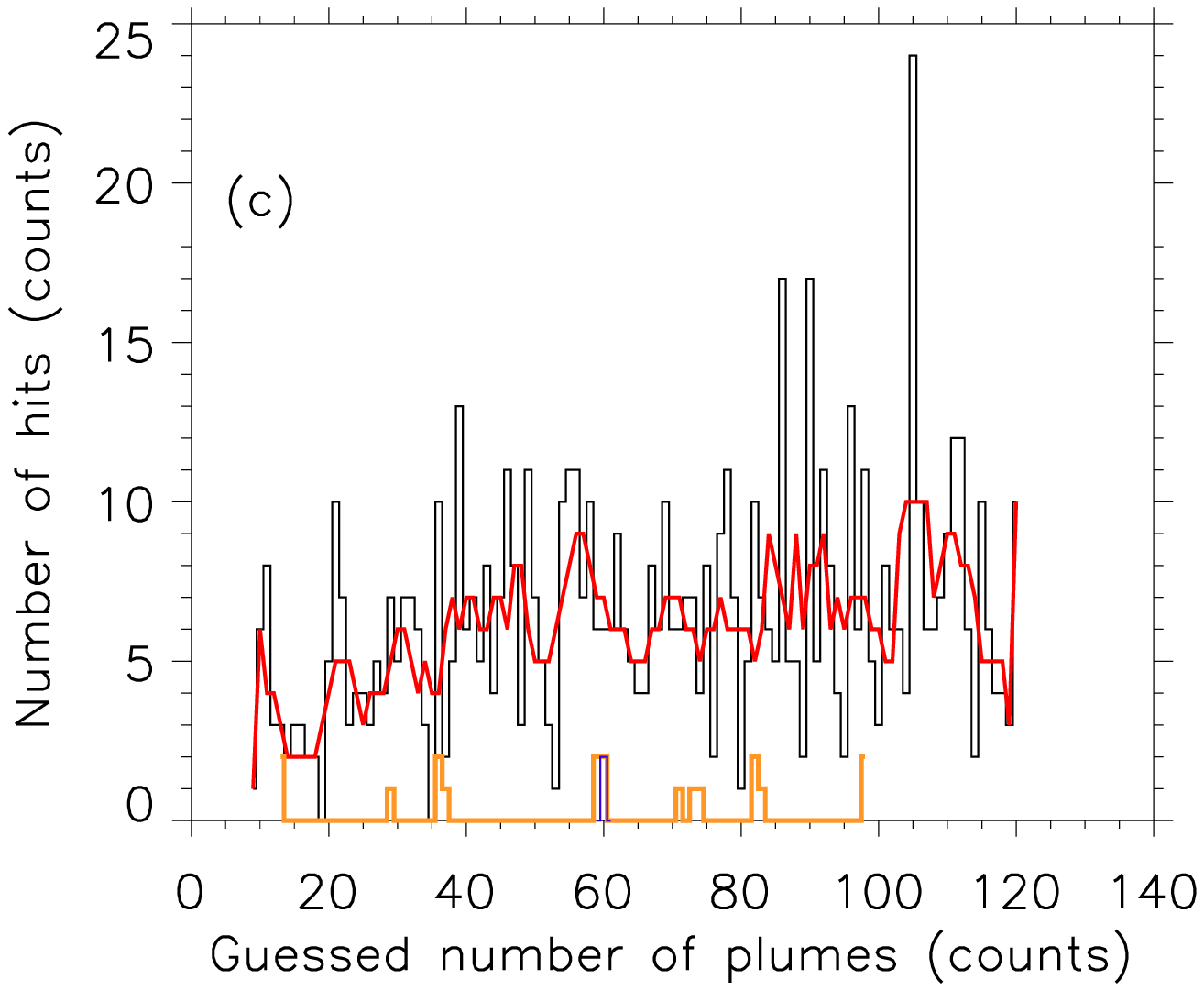}
\includegraphics[clip,trim=0cm 0cm 0cm 0cm,width=0.48\textwidth]{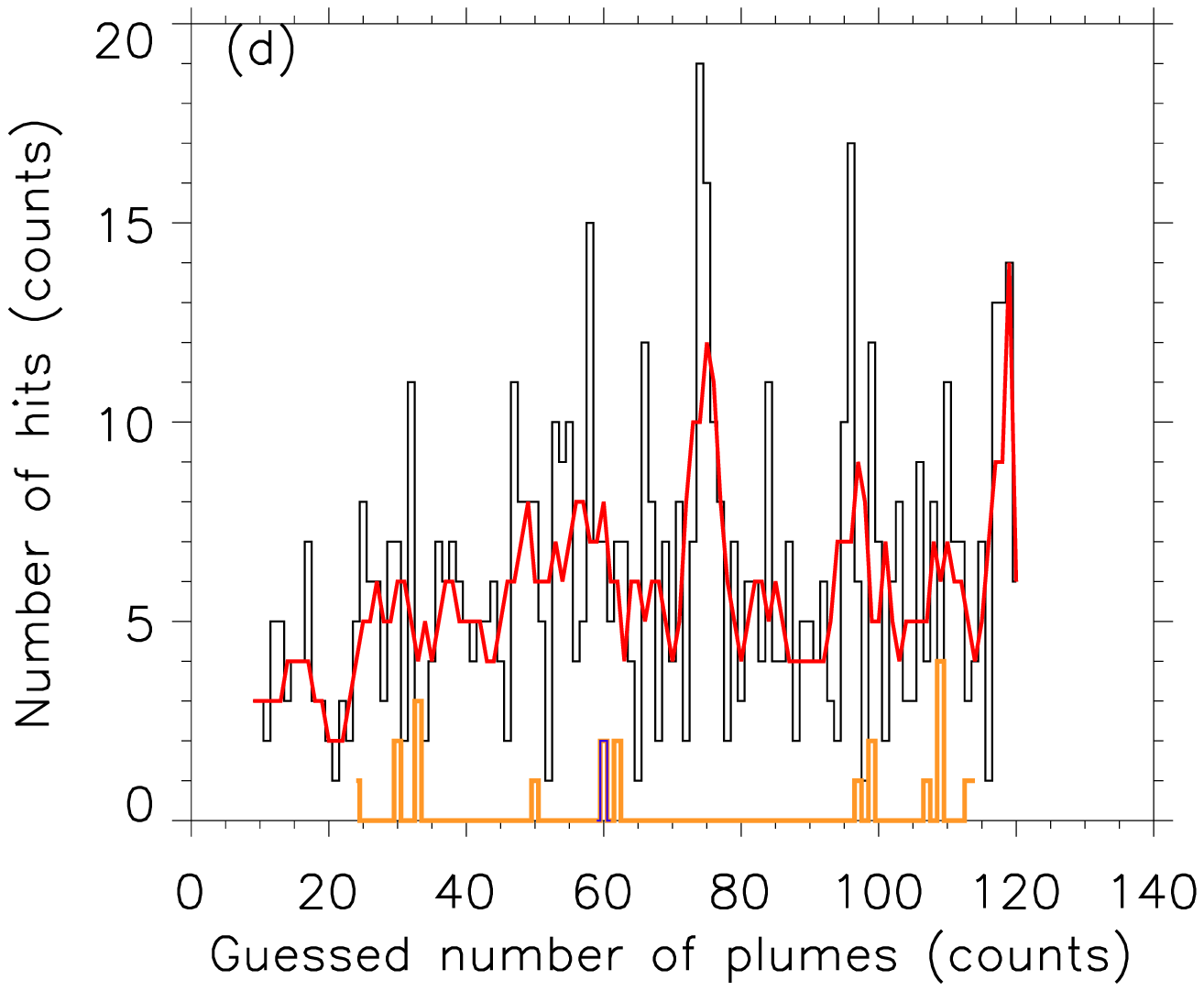}
\caption{
Histograms of number of hits at various guessed number of plumes given in the Monte Carlo simulation based on the artificial data (see Figure\,\ref{fig:lcs_sim}).
Black curves are obtained from two viewing angles and a correlation coefficient threshold of 2$\sigma$ (a: 0\degree\& 30\degree; b: 0\degree\& 90\degree; c: 0\degree\& 120\degree; d: 0\degree\& 150\degree).
Red curves are the black curves smoothed by a window of 5. 
Orange curves are obtained from three viewing angles and a correlation coefficient threshold of 2$\sigma$ (a: 0\degree, 30\degree\&60\degree; b: 0\degree, 30\degree\& 90\degree; c: 0\degree, 30\degree\& 120\degree; d: 0\degree, 30\degree\& 150\degree). 
Blue curves are obtained from three viewing angles as same as the orange curve shown in the same panel but with a correlation coefficient threshold of 3$\sigma$.}
\label{fig:ccsim}
\end{figure*}

Furthermore, here we test another case with much denser plumes, which is based on 60 CPs, each having one pixel width and locating in a region of 40\,pixels$\times$40\,pixels (40\,pixels perpendicular to line-of-sight and 40\,pixels along line-of-sight).
In Figure\,\ref{fig:lcs_sim}, we show the IVCs of the artificial data at a few viewing angles.
One can see that the peaks of the IVCs of this case are much denser than those shown in Figure\,\ref{fig:ivcs}.
We run similar Monte Carlo simulations based on these intensity variation curves and the histograms of number of hits versus guessed number of plumes are given in Figure\,\ref{fig:ccsim}.
The solutions given from two viewing angles (the black curves in Figure\,\ref{fig:ccsim}) show tendency in a few ranges of numbers (see the humps of the red curves in Figure\,\ref{fig:ccsim}), some of which are very close to the correct number.
However, the distributions are broad without dominant peaks, and thus they give little confidence for solving the correct number.
This is significantly different from the results given in the analyses of the case shown in the appendix and the real observations.
The main reason is that the plumes as shown in the IVCs of this case are much denser than those shown in the appendix and the real observations.
This indicates that high resolution and high sensitivity data will also help. 

\par
While including observations from an additional viewing angle (the orange curves in Figure\,\ref{fig:ccsim}), the distributions of hits are sparse, and thus allow a much better chance for getting the correct answer.
Furthermore, while the correlation coefficient threshold is 3$\sigma$, we found the Monte Carlo simulations return exactly the right number (see the blue curves in Figure\,\ref{fig:ccsim}) even though the plumes are relatively dense in the region.
This indicates that observations from three different viewing angles can actually solve the problem, but the correlation coefficient threshold has to be carefully selected.
Especially in analyses of real observations, it is even more tricky for selection of the threshold.
(For example, if a threshold of $3\sigma$ is used, no hit will be found in the cases shown in Section\,\ref{sect_method}. )
Therefore, to observe the Sun simultaneously from more than two viewing angles like that proposed in the concept of the Solar Ring mission will help distinguish the plasma structures in the corona and the solar wind, even though the structures are quite dense.

\section{Conclusion}
\label{sect_con}
Coronal plumes (CPs) are one of the common structures rooting in the solar coronal holes where the solar winds are originated. 
CPs have been thought to contribute significantly to the origin of the solar wind, but that remains under debate.
The population of CPs in coronal holes is crucial in understanding their significance.
Using the observations from the twin satellites of STEREO and Monte Carlo simulations, here we make the first attempt to estimate the population of CPs above the solar polar coronal holes.
We found that CPs occupy about 2--3.4\% of the area of the polar coronal holes near the solar minimum.
We estimate that CPs contribute about 9--15\% of the solar wind originating in the polar coronal hole.
These results could be affected by a few factors, including the resolution and sensitivity of the instruments and relevant assumptions used in the method (e.g. the identification of peaks, the assumption of the identical emission and size of each plume thread and the depth along line-of-sight).
Nevertheless, our results indicate that bright plume threads are not the major (in volume) structures of the polar coronal holes
and our attempts show that stereoscopic observations could be powerful in resolving structures in the corona.

\par
Our tests using artificial data found that the method can well work out the number of CPs with observations from more than two viewing angles even in the cases that CPs are relatively dense.
This demands simultaneous observations of the Sun from identical instruments at various viewing angles as described in the recently proposed Solar Ring mission\,\citep{2020ScChE..63.1699W}.

\appendix  
\section{ A Test to The Method}
\begin{figure*}
\centering
\includegraphics[clip,trim=1.5cm 0.5cm 0.2cm 0.2cm,width=\linewidth]{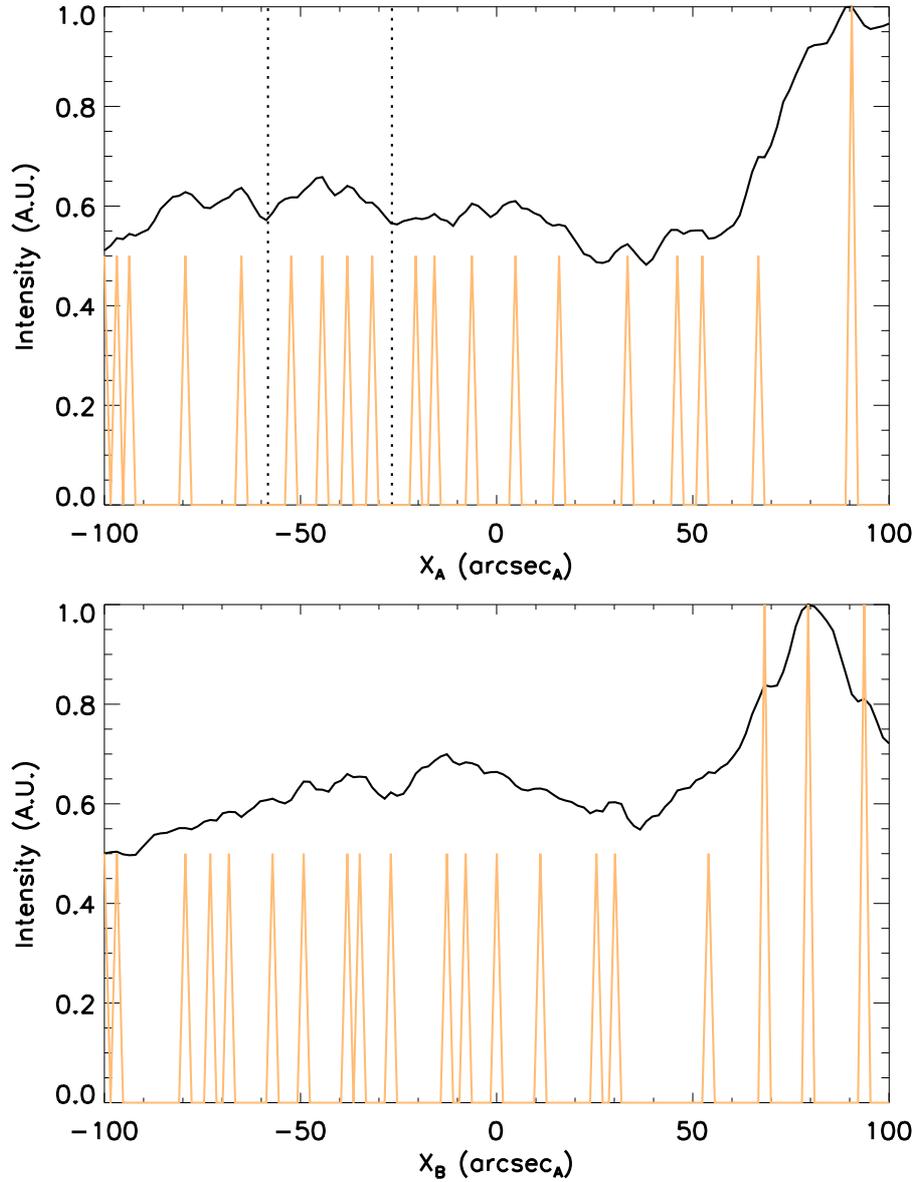}
\caption{
Examples of intensity variation curves (IVCs) obtained from the STEREO-A (top panel) and STEREO-B (bottom panel) on 2007 September 10 while {the two satellites} are separated by about 30\degree.
The arcsec units of the X-axis are both scaled to that of STEREO-A.
The black lines are the normalised original IVCs and the orange lines are the normalised clean curves generated from the identified local peaks (see the main text for details).
The dotted lines in the top panel denote the range of X-axis where four plume threads locate next to each other apparently.
Both the original and clean curves have been normalised to 1 in order to compare them.
}
\label{fig:ivcs}
\end{figure*}

\begin{figure*}
\centering
\includegraphics[clip,trim=0cm 0.3cm 0cm 1cm,width=\linewidth]{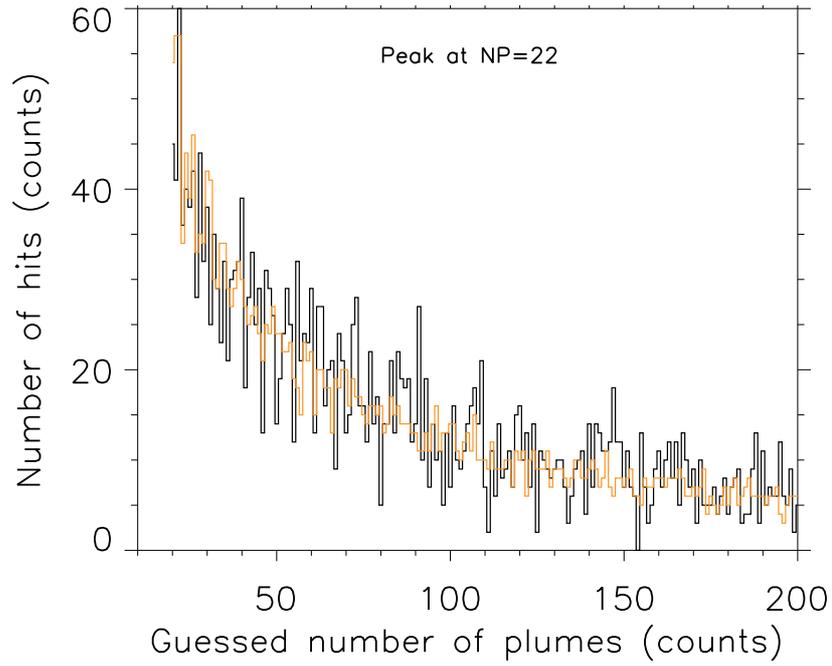}
\caption{
Histograms of number of hits at various guessed number of plumes given in the Monte Carlo simulation based on the observations shown in Figure\,\ref{fig:ivcs}, i.e. the two satellites are separated by 30\degree.
The histogram peaks at the point that the number of plumes is 22.
The black curves are the histogram based on each guessed plume number being given $10^4$ random combinations of locations, while the orange curves are based on $10^5$ random combinations and the number of hits are divided by 10.
}
\label{fig:cchist030}
\end{figure*}

\begin{figure*}
\centering
\includegraphics[clip,trim=1cm 0cm 0cm 0cm,width=0.48\textwidth]{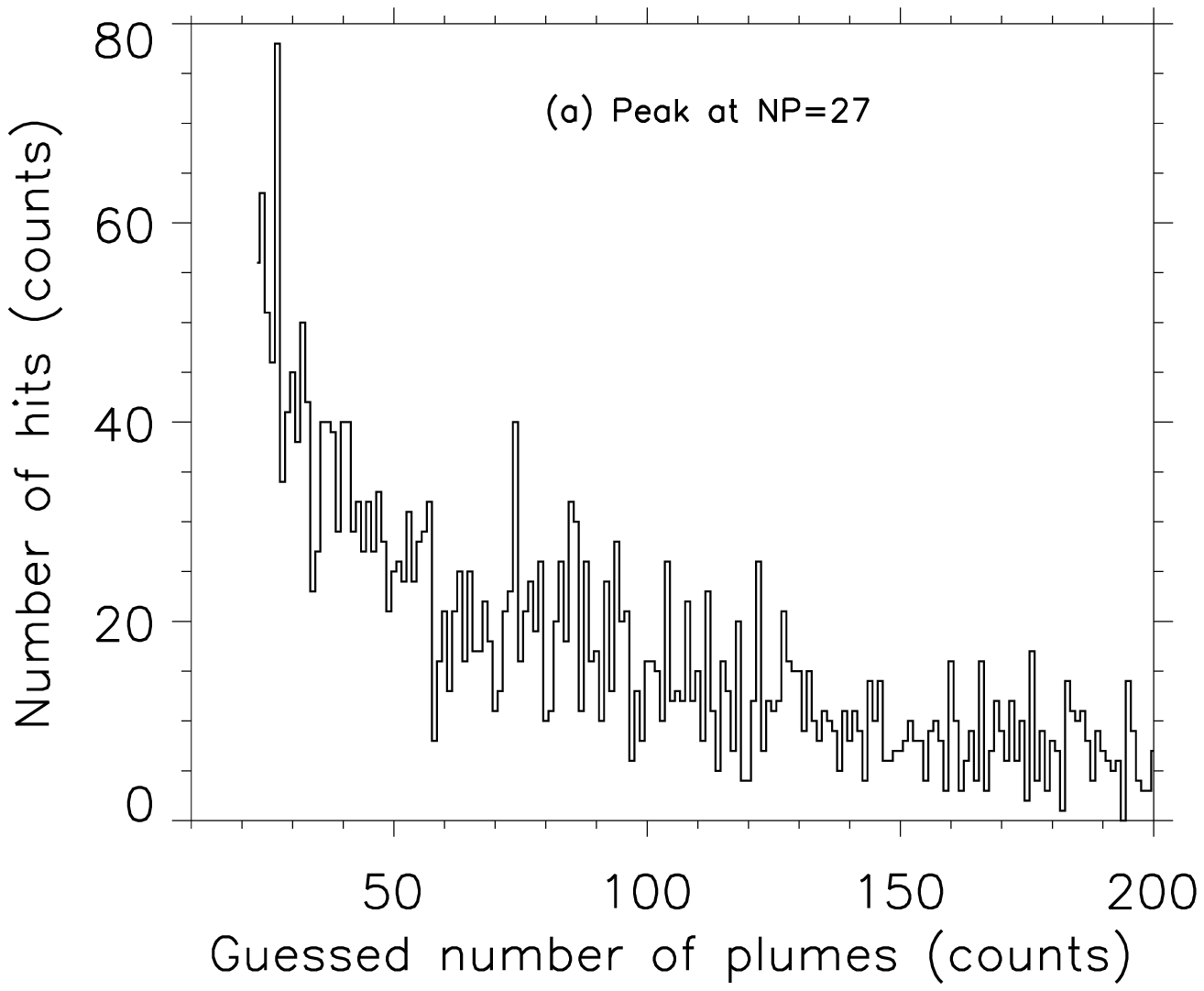}
\includegraphics[clip,trim=1cm 0cm 0cm 0cm,width=0.48\textwidth]{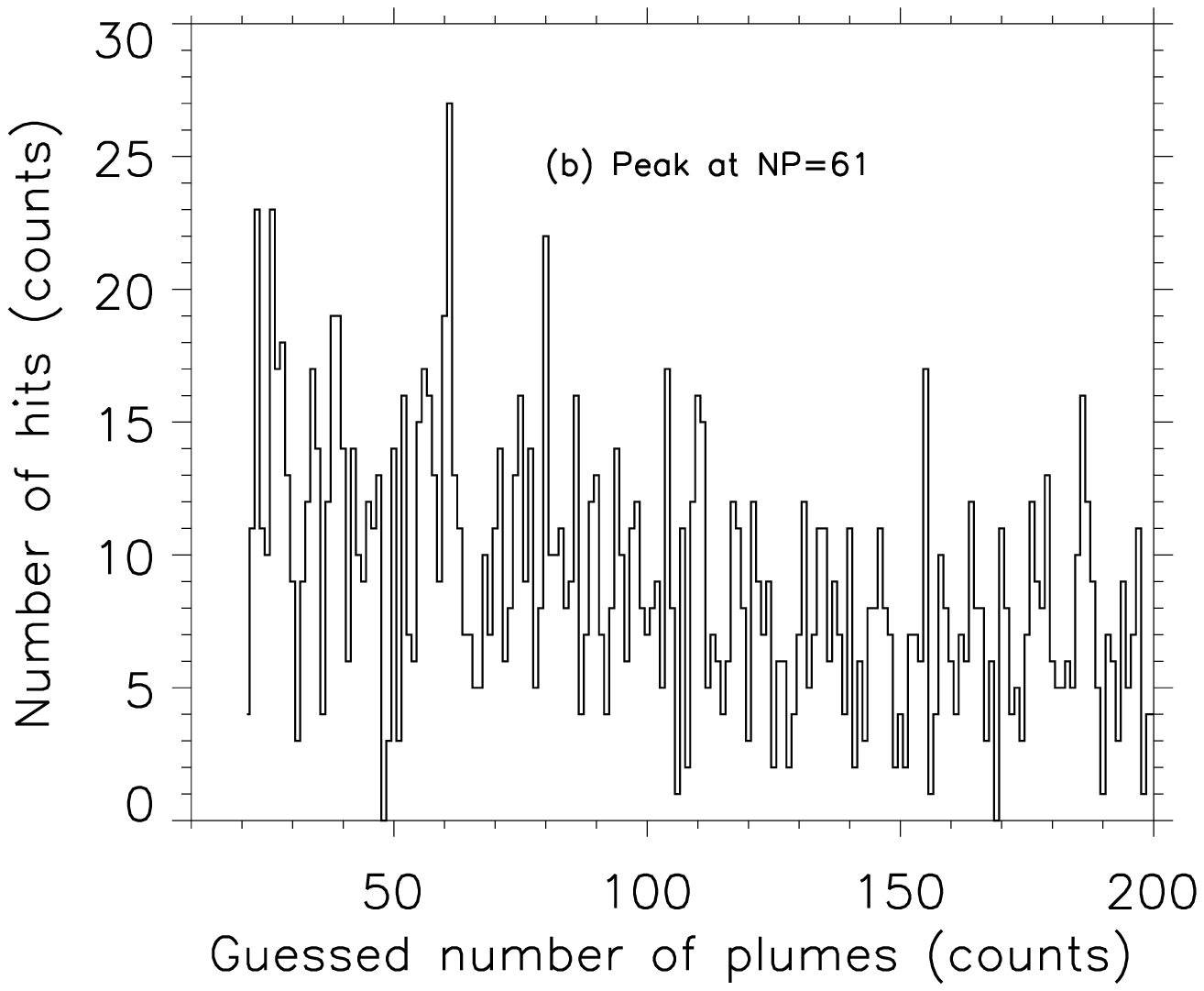}\\
\includegraphics[clip,trim=1cm 0cm 0cm 0cm,width=0.48\textwidth]{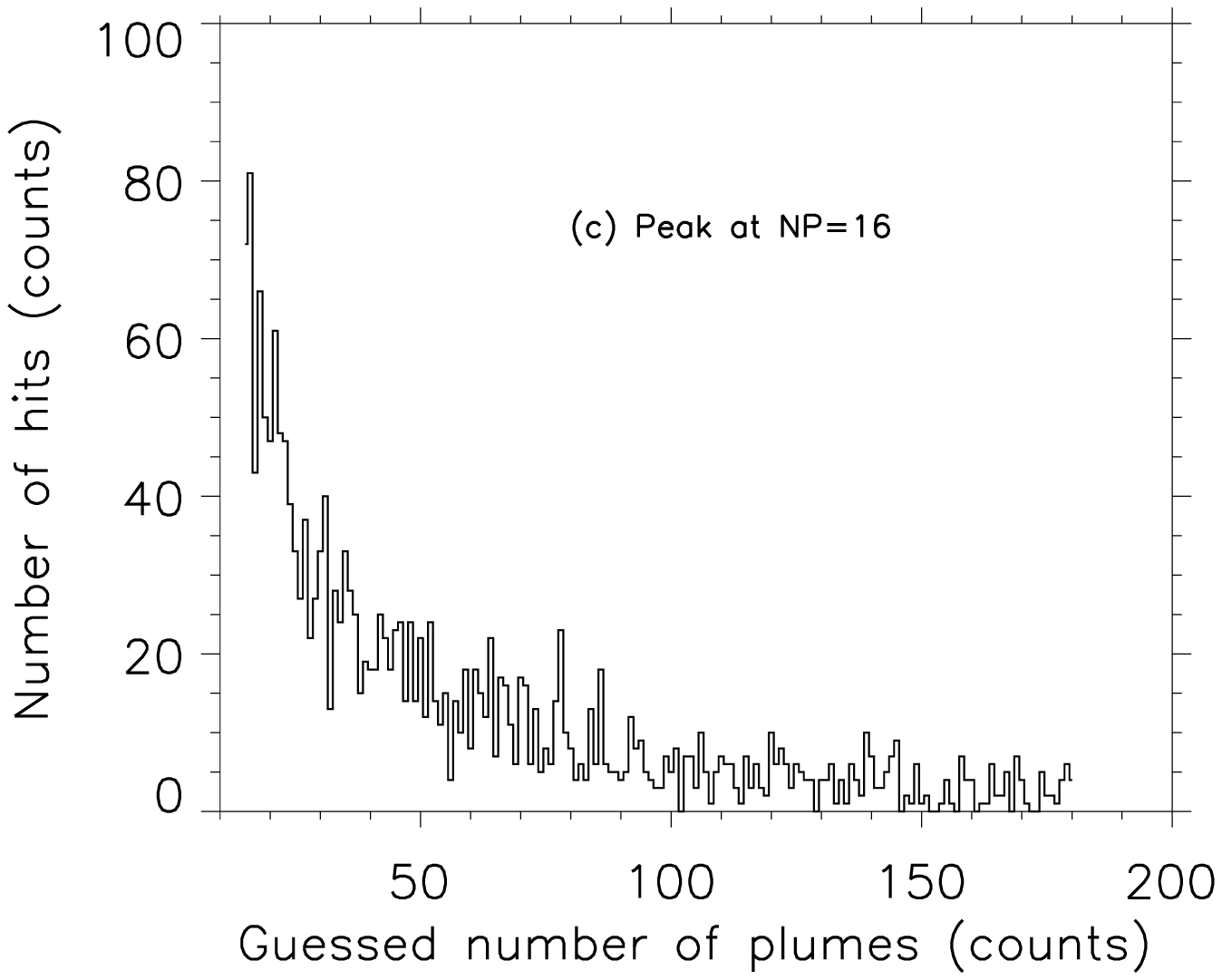}
\includegraphics[clip,trim=1cm 0cm 0cm 0cm,width=0.48\textwidth]{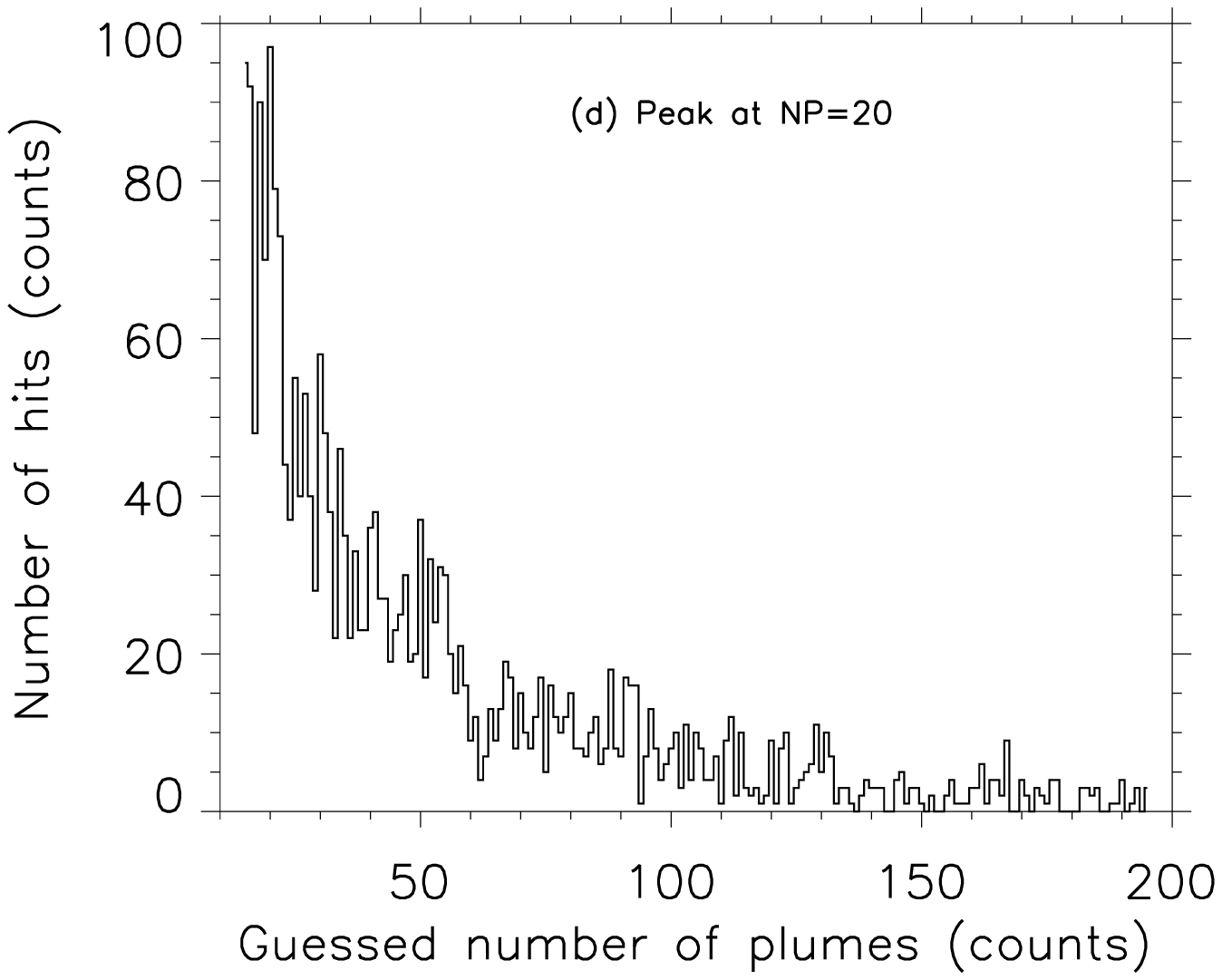}
\caption{
The same as Figure\,\ref{fig:cchist030}, but for the observations while the two satellites are separated by 60\degree\ (a), 90\degree\ (b), 120\degree\ (c) and 150\degree\ (d).
}
\label{fig:cchist}
\end{figure*}

\begin{figure*}
\centering
\includegraphics[clip,trim=1cm 1cm .3cm .3cm,width=\textwidth]{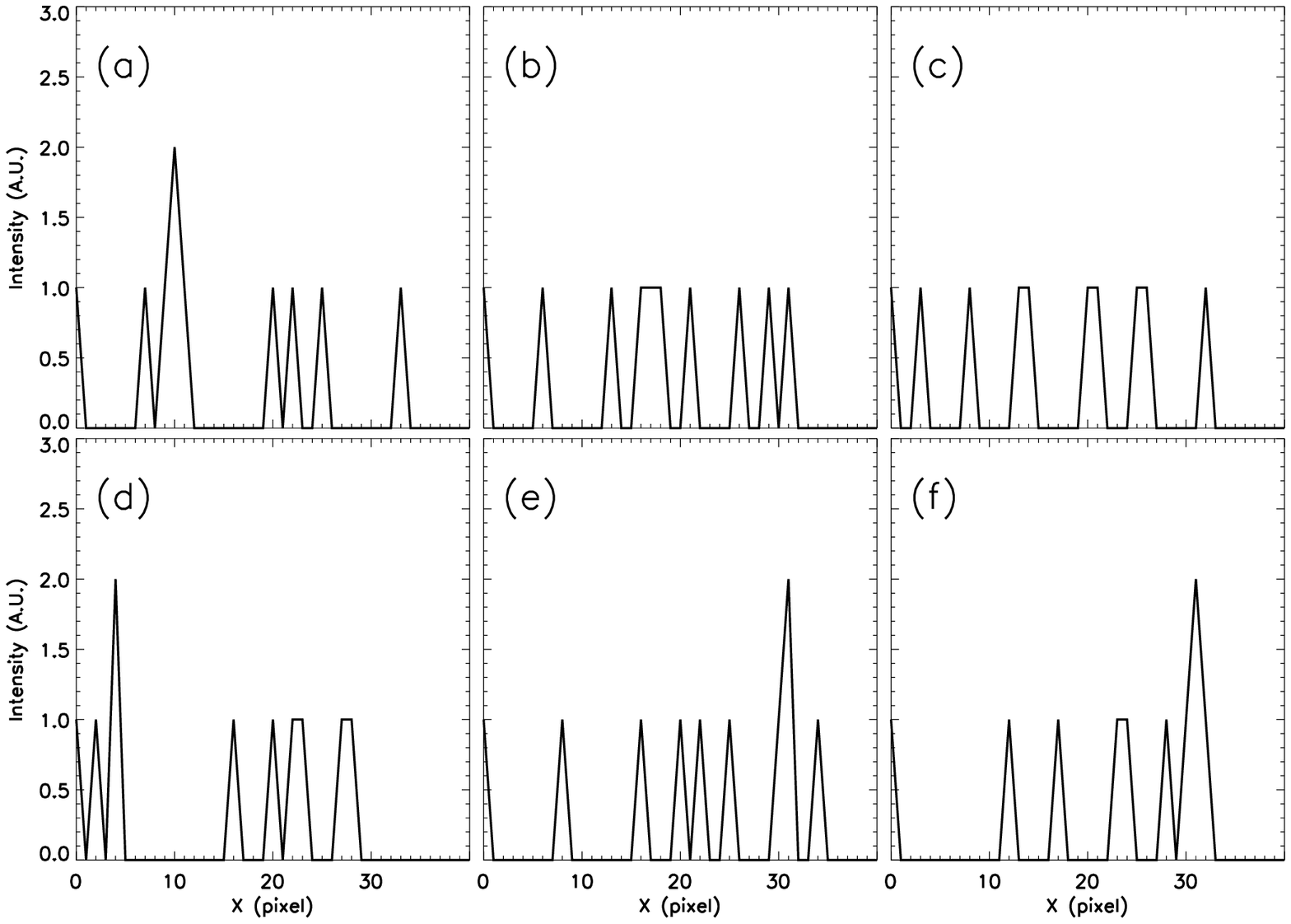}
\caption{The intensity variation curves of the artificial data with 10 plumes in a region of 40\,pixels$\times$ 40\,pixels  viewed at 0\degree\ (a), 30\degree\ (b), 60\degree\ (c), 90\degree\ (d), 120\degree\ (e) and 150\degree\ (f).
}
\label{fig:lcs_test}
\end{figure*}

\begin{figure*}
\centering
\includegraphics[clip,trim=0cm 0cm 0cm 0cm,width=0.32\textwidth]{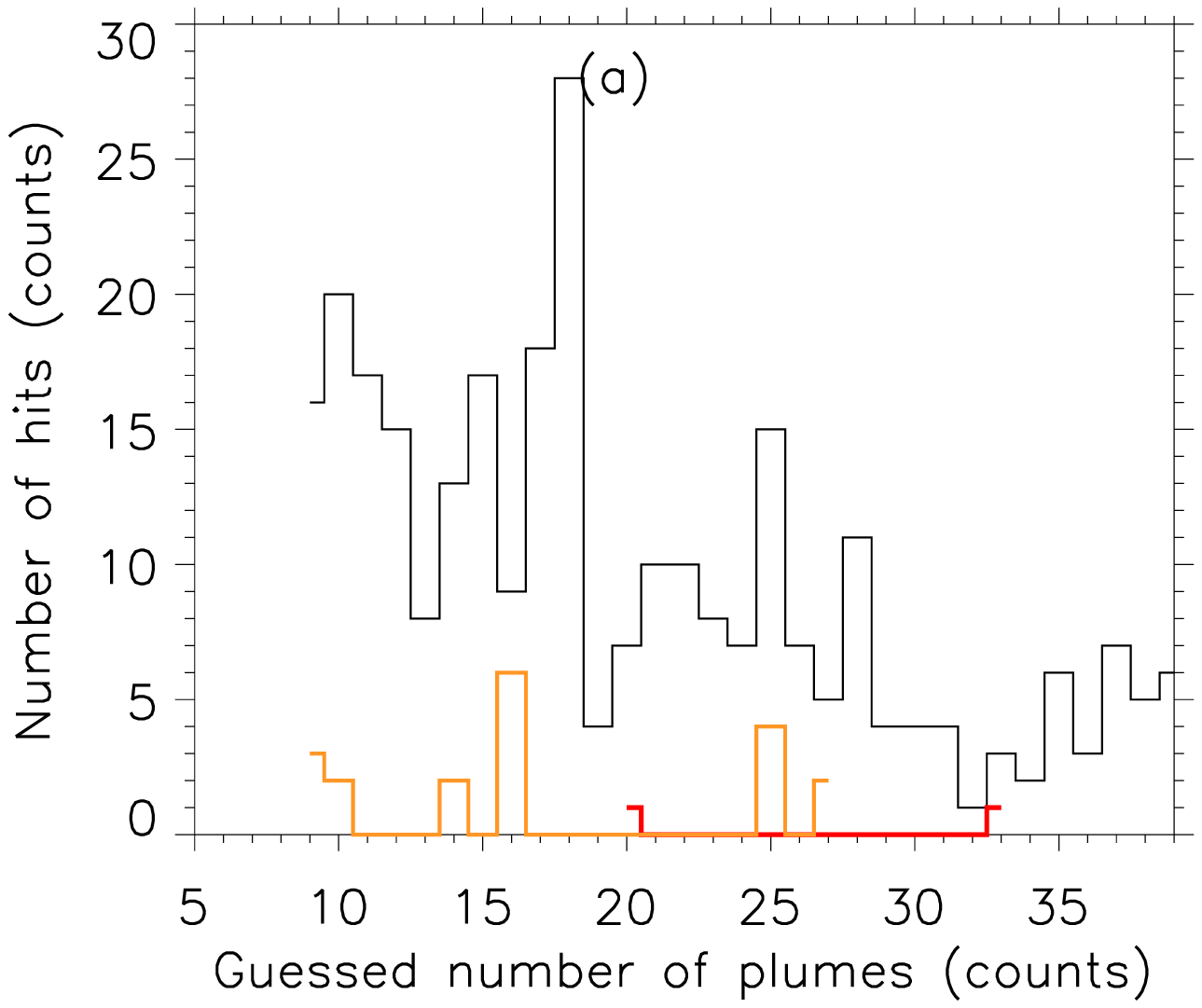}
\includegraphics[clip,trim=0cm 0cm 0cm 0cm,width=0.32\textwidth]{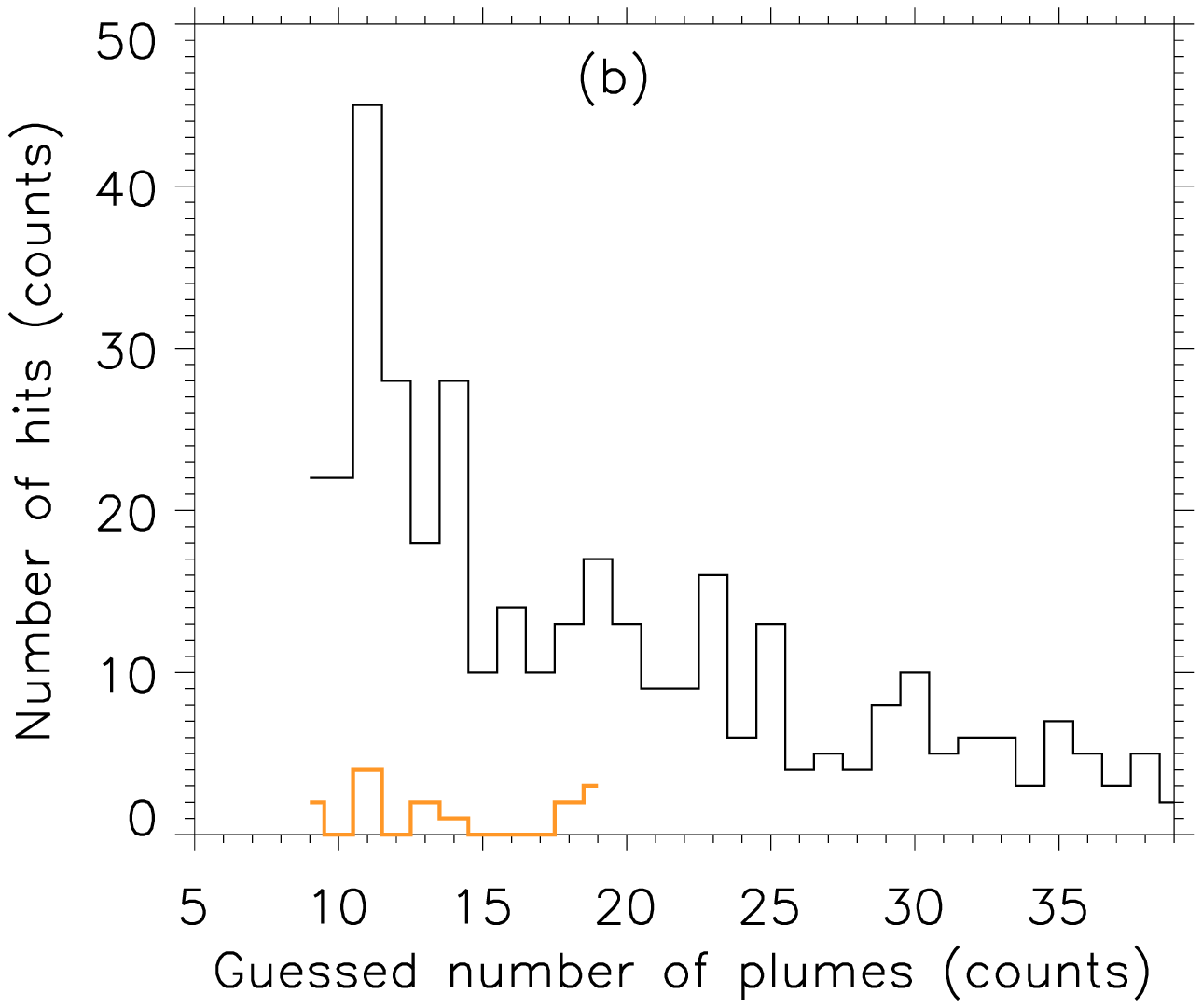}
\includegraphics[clip,trim=0cm 0cm 0cm 0cm,width=0.32\textwidth]{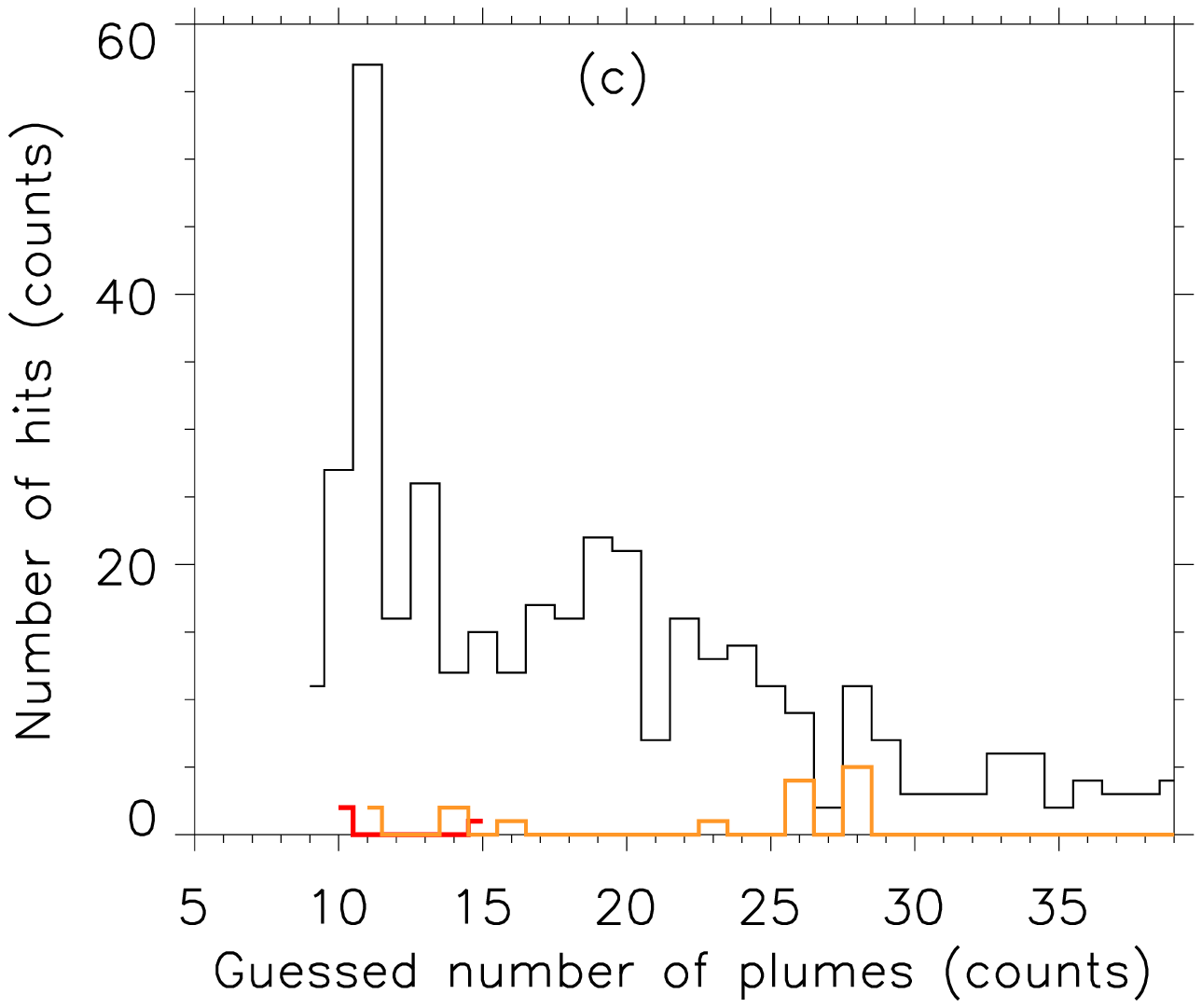}\\
\includegraphics[clip,trim=0cm 0cm 0cm 0cm,width=0.32\textwidth]{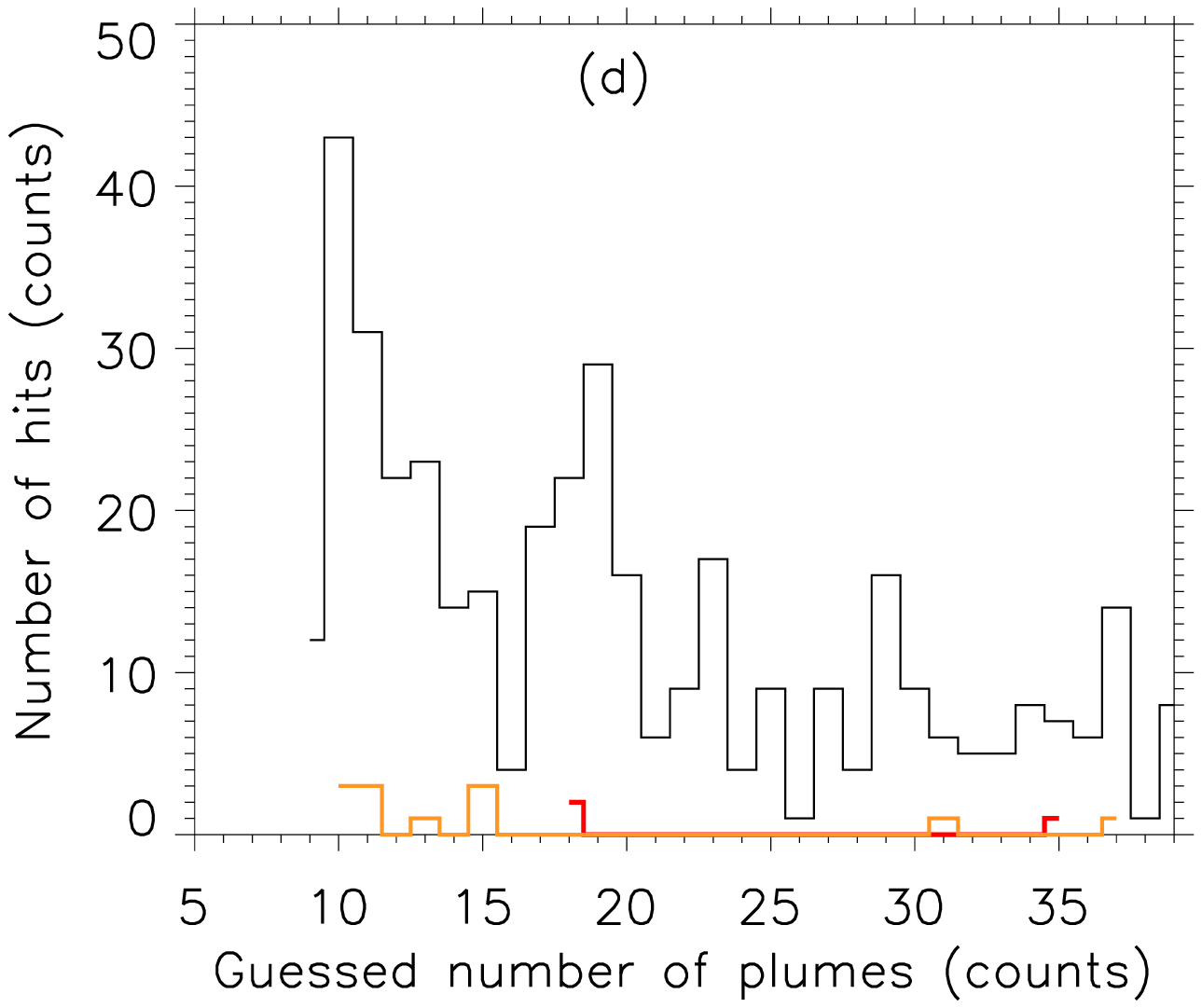}
\includegraphics[clip,trim=0cm 0cm 0cm 0cm,width=0.32\textwidth]{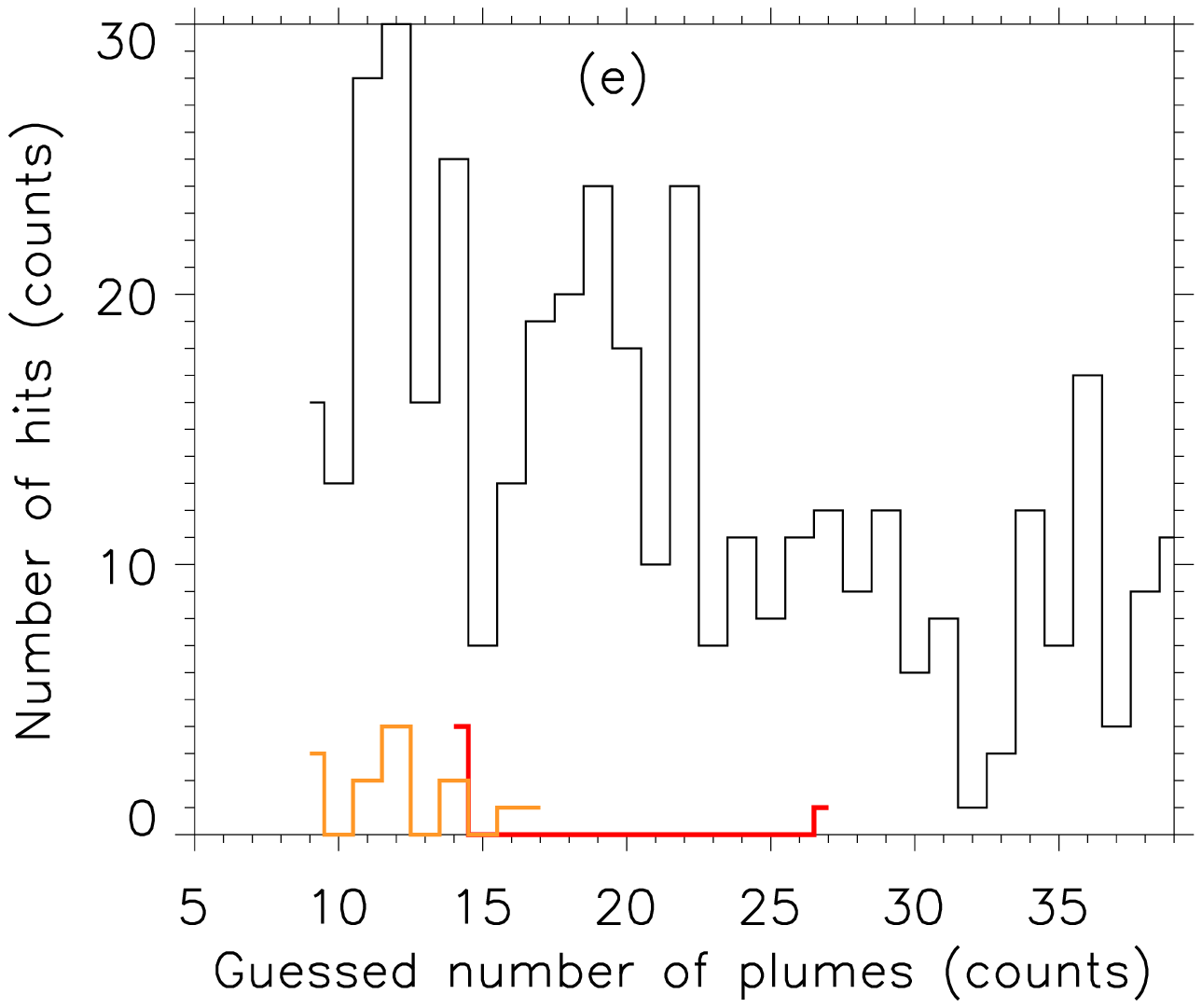}
\includegraphics[clip,trim=0cm 0cm 0cm 0cm,width=0.32\textwidth]{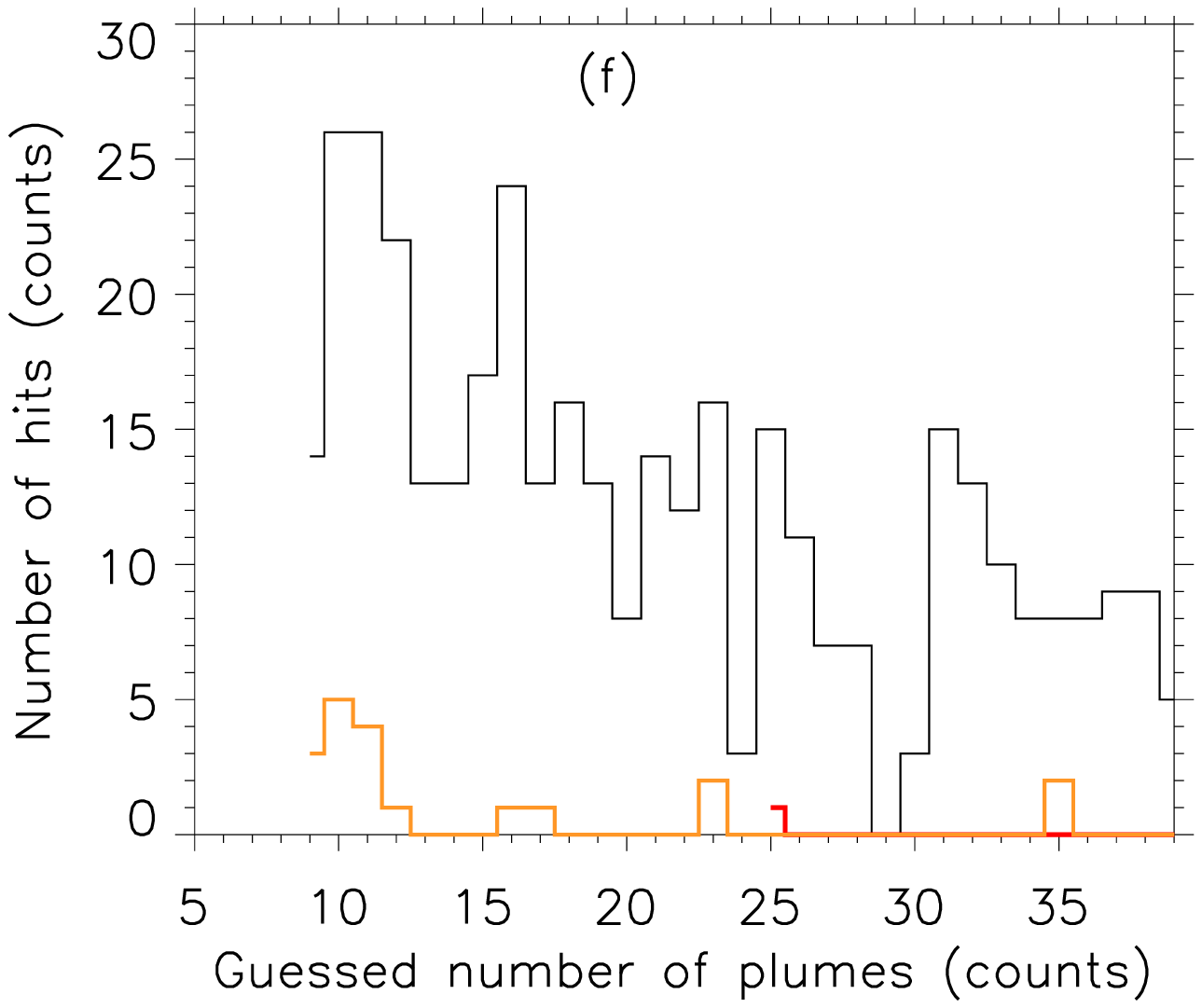}
\caption{
Histograms of number of hits at various guessed number of plumes given in the Monte Carlo simulation based on the artificial data with 10 plumes (see Figure\,\ref{fig:lcs_test}).
Black curves are obtained from two viewing angles and a correlation coefficient threshold of 2$\sigma$ (a: 0\degree\& 30\degree; b: 0\degree\& 60\degree; c: 0\degree\& 90\degree; d: 0\degree\& 120\degree; e: 0\degree\& 150\degree; f: 60\degree\& 120\degree).
Red curves are based on two viewing angles as same as those used in the black curves but with a correlation coefficient threshold of 3$\sigma$. (Red curves are not shown in panel b because none has satisfied the threshold.) 
Orange curves are obtained from three viewing angles and a correlation coefficient threshold of 2$\sigma$ (a: 0\degree, 30\degree\&60\degree; b: 0\degree, 60\degree\& 90\degree; c: 0\degree, 30\degree\& 90\degree; d: 0\degree, 30\degree\& 120\degree; e: 0\degree, 30\degree\& 150\degree; f: 0\degree, 60\degree\& 120\degree).}
\label{fig:cctest}
\end{figure*}

In order to test the method as described in Section\,\ref{sect_method}, we apply it on a set of artificial data based on that 10 plumes (each has a size of a pixel)  root in an area of 40\,pixels$\times$ 40\,pixels.
In Figure\,\ref{fig:lcs_test}, we show the IVCs at a set of viewing angles.
We can see that in this case the local peaks in the IVCs are sparse, and this is similar to those determined from observations (as shown in Figure\,\ref{fig:ivcs}).
In Figure\,\ref{fig:cctest}, we show the histograms of the distributions of the correlation coefficients derived from the Monte Carlo simulations of these artificial data.
We can see that the distributions based on two viewing angles (black curves in Figure\,\ref{fig:cctest}) show clear dominant peaks near the right number.
The numbers of plumes indicated by these dominant peaks are from 10 to 12, except the case based on 0\degree\ and 30\degree.
Although the histogram of the case of  0\degree\ and 30\degree\ has dominant peak at the guessed plume number of 18, the distribution around the plume number of 10 still appears to be larger than the rest in general.
Therefore, one can have a good estimation of the number of plumes based on the histograms of the number of hits based on two viewing angles within an error of about 20\%.
If we increase the correlation coefficient threshold to $3\sigma$, we can see that very few (even none) hits can be obtained (see the red curves in Figure\,\ref{fig:cctest}).
It appears that one cannot get a better estimation of the plume number based on this threshold than that from $2\sigma$.

\par
While we include additional data from a third viewing angle, we found that the histograms of the number of hits using a threshold of $2\sigma$ distribute at a few isolated numbers including one really close to the correct one (see the orange curves in Figure\,\ref{fig:cctest}). 
Although this does not work out the exact answer, it gives a much better chance if one has to give a guess.
If we are increasing the threshold further, the data from three viewing angles can derive a unique number that is very close to the answer  at some point.
However, we fail to find a universal threshold for all, and that seems to be different from case to case.

\begin{acks}
The authors thank the anonymous referee for the constructive comments that have helped improve a lot the paper.
This research is supported by National Natural Science Foundation of China (41974201,U1831112), the Strategic Priority Program of CAS (XDA15017300) and the Young Scholar Program of Shandong University, Weihai (2017WHWLJH07).
STEREO is a project of NASA and we are grateful to the instrument team for collecting and preparing the data.

~\\
{\bf Disclosure of Potential Conflicts of Interest} ~The authors declare that they have no conflicts of interest.
\end{acks}

\bibliographystyle{spr-mp-sola}
\bibliography{plume}

\begin{thebibliography}{80}
\ifx\bisbn     \undefined \def\bisbn  #1{ISBN #1}\fi
\ifx\binits    \undefined \def\binits#1{#1}\fi
\ifx\bauthor   \undefined \def\bauthor#1{#1}\fi
\ifx\batitle   \undefined \def\batitle#1{#1}\fi
\ifx\bjtitle   \undefined \def\bjtitle#1{\textit{#1}}\fi
\ifx\bvolume   \undefined \def\bvolume#1{\textbf{#1}}\fi
\ifx\byear     \undefined \def\byear#1{#1}\fi
\ifx\bissue    \undefined \def\bissue#1{#1}\fi
\ifx\bfpage    \undefined \def\bfpage#1{#1}\fi
\ifx\blpage    \undefined \def\blpage #1{#1}\fi
\ifx\burl      \undefined \def\burl#1{#1}\fi
\ifx\href      \undefined \def\href#1#2{#2}\fi
\ifx\betal     \undefined \def\betal{et al.}\fi
\ifx\bctitle   \undefined \def\bctitle#1{#1}\fi
\ifx\beditor   \undefined \def\beditor#1{#1}\fi
\ifx\bbtitle   \undefined \def\bbtitle#1{\textit{#1}}\fi
\ifx\bedition  \undefined \def\bedition#1{#1}\fi
\ifx\bseriesno \undefined \def\bseriesno#1{\textbf{#1}}\fi
\ifx\blocation \undefined \def\blocation#1{#1}\fi
\ifx\bsertitle \undefined \def\bsertitle#1{\textit{#1}}\fi
\ifx\bsnm      \undefined \def\bsnm#1{#1}\fi
\ifx\bsuffix   \undefined \def\bsuffix#1{#1}\fi
\ifx\bparticle \undefined \def\bparticle#1{#1}\fi
\ifx\barticle  \undefined \def\barticle#1{}\fi
\ifx\binstitute  \undefined \def\binstitute#1{#1}\fi
\ifx\bpublisher  \undefined \def\bpublisher#1{#1}\fi
\ifx\doiurl    \undefined \def\doiurl#1{\href{#1}{DOI}}\fi
\makeatletter
\def\safeHref#1#2#3{\in@{http}{#2}\ifin@\href{#2}{#3}\else\href{#1#2}{#3}\fi}
\makeatother
\ifx\adsurl    \undefined
  \def\adsurl#1{\safeHref{https://ui.adsabs.harvard.edu/abs/}{#1}{ADS}}\fi
\ifx\arxivurl  \undefined
  \def\arxivurl#1{\safeHref{http://arxiv.org/abs/}{#1}{arXiv}}\fi
\ifx\botherref \undefined \def\botherref#1{}\fi
\ifx\url       \undefined \def\url#1{#1}\fi
\ifx\bchapter  \undefined \def\bchapter#1{}\fi
\ifx\bbook     \undefined \def\bbook#1{}\fi
\ifx\bcomment  \undefined \def\bcomment#1{#1}\fi
\ifx\oauthor   \undefined \def\oauthor#1{#1}\fi
\ifx\citeauthoryear \undefined\def \citeauthoryear#1{#1}\fi
\def\endbibitem {}
\ifx\bconflocation  \undefined \def\bconflocation#1{#1} \fi

\bibitem[\protect\citeauthoryear{{Ahmad} and
  {Withbroe}}{1977}]{1977SoPh...53..397A}
\begin{barticle}
\bauthor{\bsnm{{Ahmad}}, \binits{I.A.}},
\bauthor{\bsnm{{Withbroe}}, \binits{G.L.}}:
\byear{1977},
\batitle{{EUV analysis of polar plumes.}}
\bjtitle{\solphys}
\bvolume{53},
\bfpage{397}.
\doiurl{https://doi.org/10.1007/BF00160283}.
\adsurl{1977SoPh...53..397A}.
\end{barticle}
\endbibitem

\bibitem[\protect\citeauthoryear{{Aschwanden}}{2011}]{2011LRSP....8....5A}
\begin{barticle}
\bauthor{\bsnm{{Aschwanden}}, \binits{M.J.}}:
\byear{2011},
\batitle{{Solar Stereoscopy and Tomography}}.
\bjtitle{Living Reviews in Solar Physics}
\bvolume{8},
\bfpage{5}.
\doiurl{https://doi.org/10.12942/lrsp-2011-5}.
\adsurl{2011LRSP....8....5A}.
\end{barticle}
\endbibitem

\bibitem[\protect\citeauthoryear{{Aschwanden}}{2013}]{2013ApJ...763..115A}
\begin{barticle}
\bauthor{\bsnm{{Aschwanden}}, \binits{M.J.}}:
\byear{2013},
\batitle{{Nonlinear Force-free Magnetic Field Fitting to Coronal Loops with and
  without Stereoscopy}}.
\bjtitle{\apj}
\bvolume{763},
\bfpage{115}.
\doiurl{https://doi.org/10.1088/0004-637X/763/2/115}.
\adsurl{2013ApJ...763..115A}.
\end{barticle}
\endbibitem

\bibitem[\protect\citeauthoryear{{Aschwanden}
  et~al.}{2009}]{2009ApJ...695...12A}
\begin{barticle}
\bauthor{\bsnm{{Aschwanden}}, \binits{M.J.}},
\bauthor{\bsnm{{Wuelser}}, \binits{J.-P.}},
\bauthor{\bsnm{{Nitta}}, \binits{N.V.}},
\bauthor{\bsnm{{Lemen}}, \binits{J.R.}},
\bauthor{\bsnm{{Sandman}}, \binits{A.}}:
\byear{2009},
\batitle{{First Three-Dimensional Reconstructions of Coronal Loops with the
  STEREO A+B Spacecraft. III. Instant Stereoscopic Tomography of Active
  Regions}}.
\bjtitle{\apj}
\bvolume{695},
\bfpage{12}.
\doiurl{https://doi.org/10.1088/0004-637X/695/1/12}.
\adsurl{2009ApJ...695...12A}.
\end{barticle}
\endbibitem

\bibitem[\protect\citeauthoryear{{Avallone} et~al.}{2018}]{2018ApJ...861..111A}
\begin{barticle}
\bauthor{\bsnm{{Avallone}}, \binits{E.A.}},
\bauthor{\bsnm{{Tiwari}}, \binits{S.K.}},
\bauthor{\bsnm{{Panesar}}, \binits{N.K.}},
\bauthor{\bsnm{{Moore}}, \binits{R.L.}},
\bauthor{\bsnm{{Winebarger}}, \binits{A.}}:
\byear{2018},
\batitle{{Critical Magnetic Field Strengths for Solar Coronal Plumes in Quiet
  Regions and Coronal Holes?}}
\bjtitle{\apj}
\bvolume{861},
\bfpage{111}.
\doiurl{https://doi.org/10.3847/1538-4357/aac82c}.
\adsurl{2018ApJ...861..111A}.
\end{barticle}
\endbibitem

\bibitem[\protect\citeauthoryear{{Brooks} et~al.}{2020}]{2020ApJ...894..144B}
\begin{barticle}
\bauthor{\bsnm{{Brooks}}, \binits{D.H.}},
\bauthor{\bsnm{{Winebarger}}, \binits{A.R.}},
\bauthor{\bsnm{{Savage}}, \binits{S.}},
\bauthor{\bsnm{{Warren}}, \binits{H.P.}},
\bauthor{\bsnm{{De Pontieu}}, \binits{B.}},
\bauthor{\bsnm{{Peter}}, \binits{H.}},
\bauthor{\bsnm{{Cirtain}}, \binits{J.W.}},
\bauthor{\bsnm{{Golub}}, \binits{L.}},
\bauthor{\bsnm{{Kobayashi}}, \binits{K.}},
\bauthor{\bsnm{{McIntosh}}, \binits{S.W.}},
\bauthor{\bsnm{{McKenzie}}, \binits{D.}},
\bauthor{\bsnm{{Morton}}, \binits{R.}},
\bauthor{\bsnm{{Rachmeler}}, \binits{L.}},
\bauthor{\bsnm{{Testa}}, \binits{P.}},
\bauthor{\bsnm{{Tiwari}}, \binits{S.}},
\bauthor{\bsnm{{Walsh}}, \binits{R.}}:
\byear{2020},
\batitle{{The Drivers of Active Region Outflows into the Slow Solar Wind}}.
\bjtitle{\apj}
\bvolume{894},
\bfpage{144}.
\doiurl{https://doi.org/10.3847/1538-4357/ab8a4c}.
\adsurl{2020ApJ...894..144B}.
\end{barticle}
\endbibitem

\bibitem[\protect\citeauthoryear{{Cheng} et~al.}{2020}]{2020ApJ...897...87C}
\begin{barticle}
\bauthor{\bsnm{{Cheng}}, \binits{L.}},
\bauthor{\bsnm{{Zhang}}, \binits{Q.}},
\bauthor{\bsnm{{Wang}}, \binits{Y.}},
\bauthor{\bsnm{{Li}}, \binits{X.}},
\bauthor{\bsnm{{Liu}}, \binits{R.}}:
\byear{2020},
\batitle{{Using Stereoscopic Observations of Cometary Plasma Tails to Infer
  Solar Wind Speed}}.
\bjtitle{\apj}
\bvolume{897},
\bfpage{87}.
\doiurl{https://doi.org/10.3847/1538-4357/ab93b6}.
\adsurl{2020ApJ...897...87C}.
\end{barticle}
\endbibitem

\bibitem[\protect\citeauthoryear{{Cho} et~al.}{2020}]{2020ApJ...900L..19C}
\begin{barticle}
\bauthor{\bsnm{{Cho}}, \binits{I.-H.}},
\bauthor{\bsnm{{Nakariakov}}, \binits{V.M.}},
\bauthor{\bsnm{{Moon}}, \binits{Y.-J.}},
\bauthor{\bsnm{{Lee}}, \binits{J.-Y.}},
\bauthor{\bsnm{{Yu}}, \binits{D.J.}},
\bauthor{\bsnm{{Cho}}, \binits{K.-S.}},
\bauthor{\bsnm{{Yurchyshyn}}, \binits{V.}},
\bauthor{\bsnm{{Lee}}, \binits{H.}}:
\byear{2020},
\batitle{{Accelerating and Supersonic Density Fluctuations in Coronal Hole
  Plumes: Signature of Nascent Solar Winds}}.
\bjtitle{\apjl}
\bvolume{900},
\bfpage{L19}.
\doiurl{https://doi.org/10.3847/2041-8213/abb020}.
\adsurl{2020ApJ...900L..19C}.
\end{barticle}
\endbibitem

\bibitem[\protect\citeauthoryear{{Cranmer}}{2009}]{2009LRSP....6....3C}
\begin{barticle}
\bauthor{\bsnm{{Cranmer}}, \binits{S.R.}}:
\byear{2009},
\batitle{{Coronal Holes}}.
\bjtitle{Living Reviews in Solar Physics}
\bvolume{6},
\bfpage{3}.
\doiurl{https://doi.org/10.12942/lrsp-2009-3}.
\adsurl{2009LRSP....6....3C}.
\end{barticle}
\endbibitem

\bibitem[\protect\citeauthoryear{{de Patoul}
  et~al.}{2013}]{2013SoPh..283..207D}
\begin{barticle}
\bauthor{\bsnm{{de Patoul}}, \binits{J.}},
\bauthor{\bsnm{{Inhester}}, \binits{B.}},
\bauthor{\bsnm{{Feng}}, \binits{L.}},
\bauthor{\bsnm{{Wiegelmann}}, \binits{T.}}:
\byear{2013},
\batitle{{2D and 3D Polar Plume Analysis from the Three Vantage Positions of
  STEREO/EUVI A, B, and SOHO/EIT}}.
\bjtitle{\solphys}
\bvolume{283},
\bfpage{207}.
\doiurl{https://doi.org/10.1007/s11207-011-9902-7}.
\adsurl{2013SoPh..283..207D}.
\end{barticle}
\endbibitem

\bibitem[\protect\citeauthoryear{{DeForest} and
  {Gurman}}{1998}]{1998ApJ...501L.217D}
\begin{barticle}
\bauthor{\bsnm{{DeForest}}, \binits{C.E.}},
\bauthor{\bsnm{{Gurman}}, \binits{J.B.}}:
\byear{1998},
\batitle{{Observation of Quasi-periodic Compressive Waves in Solar Polar
  Plumes}}.
\bjtitle{\apjl}
\bvolume{501},
\bfpage{L217}.
\doiurl{https://doi.org/10.1086/311460}.
\adsurl{1998ApJ...501L.217D}.
\end{barticle}
\endbibitem

\bibitem[\protect\citeauthoryear{{Deforest} et~al.}{1997}]{1997SoPh..175..393D}
\begin{barticle}
\bauthor{\bsnm{{Deforest}}, \binits{C.E.}},
\bauthor{\bsnm{{Hoeksema}}, \binits{J.T.}},
\bauthor{\bsnm{{Gurman}}, \binits{J.B.}},
\bauthor{\bsnm{{Thompson}}, \binits{B.J.}},
\bauthor{\bsnm{{Plunkett}}, \binits{S.P.}},
\bauthor{\bsnm{{Howard}}, \binits{R.}},
\bauthor{\bsnm{{Harrison}}, \binits{R.C.}},
\bauthor{\bsnm{{Hassler}}, \binits{D.M.}}:
\byear{1997},
\batitle{{Polar Plume Anatomy: Results of a Coordinated Observation}}.
\bjtitle{\solphys}
\bvolume{175},
\bfpage{393}.
\doiurl{https://doi.org/10.1023/A:1004955223306}.
\adsurl{1997SoPh..175..393D}.
\end{barticle}
\endbibitem

\bibitem[\protect\citeauthoryear{{DeForest} et~al.}{2007}]{2007ApJ...666..576D}
\begin{barticle}
\bauthor{\bsnm{{DeForest}}, \binits{C.E.}},
\bauthor{\bsnm{{Hagenaar}}, \binits{H.J.}},
\bauthor{\bsnm{{Lamb}}, \binits{D.A.}},
\bauthor{\bsnm{{Parnell}}, \binits{C.E.}},
\bauthor{\bsnm{{Welsch}}, \binits{B.T.}}:
\byear{2007},
\batitle{{Solar Magnetic Tracking. I. Software Comparison and Recommended
  Practices}}.
\bjtitle{\apj}
\bvolume{666},
\bfpage{576}.
\doiurl{https://doi.org/10.1086/518994}.
\adsurl{2007ApJ...666..576D}.
\end{barticle}
\endbibitem

\bibitem[\protect\citeauthoryear{{Feng}, {Inhester}, and
  {Mierla}}{2013}]{2013SoPh..282..221F}
\begin{barticle}
\bauthor{\bsnm{{Feng}}, \binits{L.}},
\bauthor{\bsnm{{Inhester}}, \binits{B.}},
\bauthor{\bsnm{{Mierla}}, \binits{M.}}:
\byear{2013},
\batitle{{Comparisons of CME Morphological Characteristics Derived from Five 3D
  Reconstruction Methods}}.
\bjtitle{\solphys}
\bvolume{282},
\bfpage{221}.
\doiurl{https://doi.org/10.1007/s11207-012-0143-1}.
\adsurl{2013SoPh..282..221F}.
\end{barticle}
\endbibitem

\bibitem[\protect\citeauthoryear{{Feng} et~al.}{2007a}]{2007ApJ...671L.205F}
\begin{barticle}
\bauthor{\bsnm{{Feng}}, \binits{L.}},
\bauthor{\bsnm{{Inhester}}, \binits{B.}},
\bauthor{\bsnm{{Solanki}}, \binits{S.K.}},
\bauthor{\bsnm{{Wiegelmann}}, \binits{T.}},
\bauthor{\bsnm{{Podlipnik}}, \binits{B.}},
\bauthor{\bsnm{{Howard}}, \binits{R.A.}},
\bauthor{\bsnm{{Wuelser}}, \binits{J.-P.}}:
\byear{2007}a,
\batitle{{First Stereoscopic Coronal Loop Reconstructions from STEREO SECCHI
  Images}}.
\bjtitle{\apjl}
\bvolume{671},
\bfpage{L205}.
\doiurl{https://doi.org/10.1086/525525}.
\adsurl{2007ApJ...671L.205F}.
\end{barticle}
\endbibitem

\bibitem[\protect\citeauthoryear{{Feng} et~al.}{2007b}]{2007SoPh..241..235F}
\begin{barticle}
\bauthor{\bsnm{{Feng}}, \binits{L.}},
\bauthor{\bsnm{{Wiegelmann}}, \binits{T.}},
\bauthor{\bsnm{{Inhester}}, \binits{B.}},
\bauthor{\bsnm{{Solanki}}, \binits{S.}},
\bauthor{\bsnm{{Gan}}, \binits{W.Q.}},
\bauthor{\bsnm{{Ruan}}, \binits{P.}}:
\byear{2007}b,
\batitle{{Magnetic Stereoscopy of Coronal Loops in NOAA 8891}}.
\bjtitle{\solphys}
\bvolume{241},
\bfpage{235}.
\doiurl{https://doi.org/10.1007/s11207-007-0370-z}.
\adsurl{2007SoPh..241..235F}.
\end{barticle}
\endbibitem

\bibitem[\protect\citeauthoryear{{Feng} et~al.}{2009}]{2009ApJ...700..292F}
\begin{barticle}
\bauthor{\bsnm{{Feng}}, \binits{L.}},
\bauthor{\bsnm{{Inhester}}, \binits{B.}},
\bauthor{\bsnm{{Solanki}}, \binits{S.K.}},
\bauthor{\bsnm{{Wilhelm}}, \binits{K.}},
\bauthor{\bsnm{{Wiegelmann}}, \binits{T.}},
\bauthor{\bsnm{{Podlipnik}}, \binits{B.}},
\bauthor{\bsnm{{Howard}}, \binits{R.A.}},
\bauthor{\bsnm{{Plunkett}}, \binits{S.P.}},
\bauthor{\bsnm{{Wuelser}}, \binits{J.P.}},
\bauthor{\bsnm{{Gan}}, \binits{W.Q.}}:
\byear{2009},
\batitle{{Stereoscopic Polar Plume Reconstructions from STEREO/SECCHI Images}}.
\bjtitle{\apj}
\bvolume{700},
\bfpage{292}.
\doiurl{https://doi.org/10.1088/0004-637X/700/1/292}.
\adsurl{2009ApJ...700..292F}.
\end{barticle}
\endbibitem

\bibitem[\protect\citeauthoryear{{Feng} et~al.}{2012}]{2012ApJ...751...18F}
\begin{barticle}
\bauthor{\bsnm{{Feng}}, \binits{L.}},
\bauthor{\bsnm{{Inhester}}, \binits{B.}},
\bauthor{\bsnm{{Wei}}, \binits{Y.}},
\bauthor{\bsnm{{Gan}}, \binits{W.Q.}},
\bauthor{\bsnm{{Zhang}}, \binits{T.L.}},
\bauthor{\bsnm{{Wang}}, \binits{M.Y.}}:
\byear{2012},
\batitle{{Morphological Evolution of a Three-dimensional Coronal Mass Ejection
  Cloud Reconstructed from Three Viewpoints}}.
\bjtitle{\apj}
\bvolume{751},
\bfpage{18}.
\doiurl{https://doi.org/10.1088/0004-637X/751/1/18}.
\adsurl{2012ApJ...751...18F}.
\end{barticle}
\endbibitem

\bibitem[\protect\citeauthoryear{{Fu} et~al.}{2014}]{2014ApJ...794..109F}
\begin{barticle}
\bauthor{\bsnm{{Fu}}, \binits{H.}},
\bauthor{\bsnm{{Xia}}, \binits{L.}},
\bauthor{\bsnm{{Li}}, \binits{B.}},
\bauthor{\bsnm{{Huang}}, \binits{Z.}},
\bauthor{\bsnm{{Jiao}}, \binits{F.}},
\bauthor{\bsnm{{Mou}}, \binits{C.}}:
\byear{2014},
\batitle{{Measurements of Outflow Velocities in on-disk Plumes from EIS/Hinode
  Observations}}.
\bjtitle{\apj}
\bvolume{794},
\bfpage{109}.
\doiurl{https://doi.org/10.1088/0004-637X/794/2/109}.
\adsurl{2014ApJ...794..109F}.
\end{barticle}
\endbibitem

\bibitem[\protect\citeauthoryear{{Gabriel}, {Bely-Dubau}, and
  {Lemaire}}{2003}]{2003ApJ...589..623G}
\begin{barticle}
\bauthor{\bsnm{{Gabriel}}, \binits{A.H.}},
\bauthor{\bsnm{{Bely-Dubau}}, \binits{F.}},
\bauthor{\bsnm{{Lemaire}}, \binits{P.}}:
\byear{2003},
\batitle{{The Contribution of Polar Plumes to the Fast Solar Wind}}.
\bjtitle{\apj}
\bvolume{589},
\bfpage{623}.
\doiurl{https://doi.org/10.1086/374416}.
\adsurl{2003ApJ...589..623G}.
\end{barticle}
\endbibitem

\bibitem[\protect\citeauthoryear{{Gabriel} et~al.}{2005}]{2005ApJ...635L.185G}
\begin{barticle}
\bauthor{\bsnm{{Gabriel}}, \binits{A.H.}},
\bauthor{\bsnm{{Abbo}}, \binits{L.}},
\bauthor{\bsnm{{Bely-Dubau}}, \binits{F.}},
\bauthor{\bsnm{{Llebaria}}, \binits{A.}},
\bauthor{\bsnm{{Antonucci}}, \binits{E.}}:
\byear{2005},
\batitle{{Solar Wind Outflow in Polar Plumes from 1.05 to 2.4 R$_{solar}$}}.
\bjtitle{\apjl}
\bvolume{635},
\bfpage{L185}.
\doiurl{https://doi.org/10.1086/499521}.
\adsurl{2005ApJ...635L.185G}.
\end{barticle}
\endbibitem

\bibitem[\protect\citeauthoryear{{Gabriel} et~al.}{2009}]{2009ApJ...700..551G}
\begin{barticle}
\bauthor{\bsnm{{Gabriel}}, \binits{A.}},
\bauthor{\bsnm{{Bely-Dubau}}, \binits{F.}},
\bauthor{\bsnm{{Tison}}, \binits{E.}},
\bauthor{\bsnm{{Wilhelm}}, \binits{K.}}:
\byear{2009},
\batitle{{The Structure and Origin of Solar Plumes: Network Plumes}}.
\bjtitle{\apj}
\bvolume{700},
\bfpage{551}.
\doiurl{https://doi.org/10.1088/0004-637X/700/1/551}.
\adsurl{2009ApJ...700..551G}.
\end{barticle}
\endbibitem

\bibitem[\protect\citeauthoryear{{Giordano} et~al.}{2000}]{2000ApJ...531L..79G}
\begin{barticle}
\bauthor{\bsnm{{Giordano}}, \binits{S.}},
\bauthor{\bsnm{{Antonucci}}, \binits{E.}},
\bauthor{\bsnm{{Noci}}, \binits{G.}},
\bauthor{\bsnm{{Romoli}}, \binits{M.}},
\bauthor{\bsnm{{Kohl}}, \binits{J.L.}}:
\byear{2000},
\batitle{{Identification of the Coronal Sources of the Fast Solar Wind}}.
\bjtitle{\apjl}
\bvolume{531},
\bfpage{L79}.
\doiurl{https://doi.org/10.1086/312525}.
\adsurl{2000ApJ...531L..79G}.
\end{barticle}
\endbibitem

\bibitem[\protect\citeauthoryear{{Gupta} et~al.}{2010}]{2010ApJ...718...11G}
\begin{barticle}
\bauthor{\bsnm{{Gupta}}, \binits{G.R.}},
\bauthor{\bsnm{{Banerjee}}, \binits{D.}},
\bauthor{\bsnm{{Teriaca}}, \binits{L.}},
\bauthor{\bsnm{{Imada}}, \binits{S.}},
\bauthor{\bsnm{{Solanki}}, \binits{S.}}:
\byear{2010},
\batitle{{Accelerating Waves in Polar Coronal Holes as Seen by EIS and SUMER}}.
\bjtitle{\apj}
\bvolume{718},
\bfpage{11}.
\doiurl{https://doi.org/10.1088/0004-637X/718/1/11}.
\adsurl{2010ApJ...718...11G}.
\end{barticle}
\endbibitem

\bibitem[\protect\citeauthoryear{{Hassler} et~al.}{1999}]{1999Sci...283..810H}
\begin{barticle}
\bauthor{\bsnm{{Hassler}}, \binits{D.M.}},
\bauthor{\bsnm{{Dammasch}}, \binits{I.E.}},
\bauthor{\bsnm{{Lemaire}}, \binits{P.}},
\bauthor{\bsnm{{Brekke}}, \binits{P.}},
\bauthor{\bsnm{{Curdt}}, \binits{W.}},
\bauthor{\bsnm{{Mason}}, \binits{H.E.}},
\bauthor{\bsnm{{Vial}}, \binits{J.-C.}},
\bauthor{\bsnm{{Wilhelm}}, \binits{K.}}:
\byear{1999},
\batitle{{Solar Wind Outflow and the Chromospheric Magnetic Network}}.
\bjtitle{Science}
\bvolume{283},
\bfpage{810}.
\doiurl{https://doi.org/10.1126/science.283.5403.810}.
\adsurl{1999Sci...283..810H}.
\end{barticle}
\endbibitem

\bibitem[\protect\citeauthoryear{{He} et~al.}{2010}]{2010AdSpR..45..303H}
\begin{barticle}
\bauthor{\bsnm{{He}}, \binits{J.-S.}},
\bauthor{\bsnm{{Tu}}, \binits{C.-Y.}},
\bauthor{\bsnm{{Tian}}, \binits{H.}},
\bauthor{\bsnm{{Marsch}}, \binits{E.}}:
\byear{2010},
\batitle{{Solar wind origins in coronal holes and in the quiet Sun}}.
\bjtitle{Advances in Space Research}
\bvolume{45},
\bfpage{303}.
\doiurl{https://doi.org/10.1016/j.asr.2009.07.020}.
\adsurl{2010AdSpR..45..303H}.
\end{barticle}
\endbibitem

\bibitem[\protect\citeauthoryear{{Howard} et~al.}{2008}]{2008SSRv..136...67H}
\begin{barticle}
\bauthor{\bsnm{{Howard}}, \binits{R.A.}},
\bauthor{\bsnm{{Moses}}, \binits{J.D.}},
\bauthor{\bsnm{{Vourlidas}}, \binits{A.}},
\bauthor{\bsnm{{Newmark}}, \binits{J.S.}},
\bauthor{\bsnm{{Socker}}, \binits{D.G.}},
\bauthor{\bsnm{{Plunkett}}, \binits{S.P.}},
\bauthor{\bsnm{{Korendyke}}, \binits{C.M.}},
\bauthor{\bsnm{{Cook}}, \binits{J.W.}},
\bauthor{\bsnm{{Hurley}}, \binits{A.}},
\bauthor{\bsnm{{Davila}}, \binits{J.M.}},
\bauthor{\bsnm{{Thompson}}, \binits{W.T.}},
\bauthor{\bsnm{{St Cyr}}, \binits{O.C.}},
\bauthor{\bsnm{{Mentzell}}, \binits{E.}},
\bauthor{\bsnm{{Mehalick}}, \binits{K.}},
\bauthor{\bsnm{{Lemen}}, \binits{J.R.}},
\bauthor{\bsnm{{Wuelser}}, \binits{J.P.}},
\bauthor{\bsnm{{Duncan}}, \binits{D.W.}},
\bauthor{\bsnm{{Tarbell}}, \binits{T.D.}},
\bauthor{\bsnm{{Wolfson}}, \binits{C.J.}},
\bauthor{\bsnm{{Moore}}, \binits{A.}},
\bauthor{\bsnm{{Harrison}}, \binits{R.A.}},
\bauthor{\bsnm{{Waltham}}, \binits{N.R.}},
\bauthor{\bsnm{{Lang}}, \binits{J.}},
\bauthor{\bsnm{{Davis}}, \binits{C.J.}},
\bauthor{\bsnm{{Eyles}}, \binits{C.J.}},
\bauthor{\bsnm{{Mapson-Menard}}, \binits{H.}},
\bauthor{\bsnm{{Simnett}}, \binits{G.M.}},
\bauthor{\bsnm{{Halain}}, \binits{J.P.}},
\bauthor{\bsnm{{Defise}}, \binits{J.M.}},
\bauthor{\bsnm{{Mazy}}, \binits{E.}},
\bauthor{\bsnm{{Rochus}}, \binits{P.}},
\bauthor{\bsnm{{Mercier}}, \binits{R.}},
\bauthor{\bsnm{{Ravet}}, \binits{M.F.}},
\bauthor{\bsnm{{Delmotte}}, \binits{F.}},
\bauthor{\bsnm{{Auchere}}, \binits{F.}},
\bauthor{\bsnm{{Delaboudiniere}}, \binits{J.P.}},
\bauthor{\bsnm{{Bothmer}}, \binits{V.}},
\bauthor{\bsnm{{Deutsch}}, \binits{W.}},
\bauthor{\bsnm{{Wang}}, \binits{D.}},
\bauthor{\bsnm{{Rich}}, \binits{N.}},
\bauthor{\bsnm{{Cooper}}, \binits{S.}},
\bauthor{\bsnm{{Stephens}}, \binits{V.}},
\bauthor{\bsnm{{Maahs}}, \binits{G.}},
\bauthor{\bsnm{{Baugh}}, \binits{R.}},
\bauthor{\bsnm{{McMullin}}, \binits{D.}},
\bauthor{\bsnm{{Carter}}, \binits{T.}}:
\byear{2008},
\batitle{{Sun Earth Connection Coronal and Heliospheric Investigation
  (SECCHI)}}.
\bjtitle{\ssr}
\bvolume{136},
\bfpage{67}.
\doiurl{https://doi.org/10.1007/s11214-008-9341-4}.
\adsurl{2008SSRv..136...67H}.
\end{barticle}
\endbibitem

\bibitem[\protect\citeauthoryear{{Huang} et~al.}{2012}]{2012A&A...548A..62H}
\begin{barticle}
\bauthor{\bsnm{{Huang}}, \binits{Z.}},
\bauthor{\bsnm{{Madjarska}}, \binits{M.S.}},
\bauthor{\bsnm{{Doyle}}, \binits{J.G.}},
\bauthor{\bsnm{{Lamb}}, \binits{D.A.}}:
\byear{2012},
\batitle{{Coronal hole boundaries at small scales. IV. SOT view. Magnetic field
  properties of small-scale transient brightenings in coronal holes}}.
\bjtitle{A\&A}
\bvolume{548},
\bfpage{A62}.
\doiurl{https://doi.org/10.1051/0004-6361/201220079}.
\adsurl{2012A&A...548A..62H}.
\end{barticle}
\endbibitem

\bibitem[\protect\citeauthoryear{{Huang} et~al.}{2017}]{2017MNRAS.464.1753H}
\begin{barticle}
\bauthor{\bsnm{{Huang}}, \binits{Z.}},
\bauthor{\bsnm{{Madjarska}}, \binits{M.S.}},
\bauthor{\bsnm{{Scullion}}, \binits{E.M.}},
\bauthor{\bsnm{{Xia}}, \binits{L.-D.}},
\bauthor{\bsnm{{Doyle}}, \binits{J.G.}},
\bauthor{\bsnm{{Ray}}, \binits{T.}}:
\byear{2017},
\batitle{{Explosive events in active region observed by IRIS and SST/CRISP}}.
\bjtitle{\mnras}
\bvolume{464},
\bfpage{1753}.
\doiurl{https://doi.org/10.1093/mnras/stw2469}.
\adsurl{2017MNRAS.464.1753H}.
\end{barticle}
\endbibitem

\bibitem[\protect\citeauthoryear{{Jiang} et~al.}{2020}]{2020ApJS..250....5J}
\begin{barticle}
\bauthor{\bsnm{{Jiang}}, \binits{H.}},
\bauthor{\bsnm{{Wang}}, \binits{J.}},
\bauthor{\bsnm{{Liu}}, \binits{C.}},
\bauthor{\bsnm{{Jing}}, \binits{J.}},
\bauthor{\bsnm{{Liu}}, \binits{H.}},
\bauthor{\bsnm{{Wang}}, \binits{J.T.L.}},
\bauthor{\bsnm{{Wang}}, \binits{H.}}:
\byear{2020},
\batitle{{Identifying and Tracking Solar Magnetic Flux Elements with Deep
  Learning}}.
\bjtitle{\apjs}
\bvolume{250},
\bfpage{5}.
\doiurl{https://doi.org/10.3847/1538-4365/aba4aa}.
\adsurl{2020ApJS..250....5J}.
\end{barticle}
\endbibitem

\bibitem[\protect\citeauthoryear{{Jiao} et~al.}{2015}]{2015ApJ...809L..17J}
\begin{barticle}
\bauthor{\bsnm{{Jiao}}, \binits{F.}},
\bauthor{\bsnm{{Xia}}, \binits{L.}},
\bauthor{\bsnm{{Li}}, \binits{B.}},
\bauthor{\bsnm{{Huang}}, \binits{Z.}},
\bauthor{\bsnm{{Li}}, \binits{X.}},
\bauthor{\bsnm{{Chandrashekhar}}, \binits{K.}},
\bauthor{\bsnm{{Mou}}, \binits{C.}},
\bauthor{\bsnm{{Fu}}, \binits{H.}}:
\byear{2015},
\batitle{{Sources of Quasi-periodic Propagating Disturbances above a Solar
  Polar Coronal Hole}}.
\bjtitle{\apjl}
\bvolume{809},
\bfpage{L17}.
\doiurl{https://doi.org/10.1088/2041-8205/809/1/L17}.
\adsurl{2015ApJ...809L..17J}.
\end{barticle}
\endbibitem

\bibitem[\protect\citeauthoryear{{Kaiser} et~al.}{2008}]{2008SSRv..136....5K}
\begin{barticle}
\bauthor{\bsnm{{Kaiser}}, \binits{M.L.}},
\bauthor{\bsnm{{Kucera}}, \binits{T.A.}},
\bauthor{\bsnm{{Davila}}, \binits{J.M.}},
\bauthor{\bsnm{{St. Cyr}}, \binits{O.C.}},
\bauthor{\bsnm{{Guhathakurta}}, \binits{M.}},
\bauthor{\bsnm{{Christian}}, \binits{E.}}:
\byear{2008},
\batitle{{The STEREO Mission: An Introduction}}.
\bjtitle{\ssr}
\bvolume{136},
\bfpage{5}.
\doiurl{https://doi.org/10.1007/s11214-007-9277-0}.
\adsurl{2008SSRv..136....5K}.
\end{barticle}
\endbibitem

\bibitem[\protect\citeauthoryear{{Krieger}, {Timothy}, and
  {Roelof}}{1973}]{1973SoPh...29..505K}
\begin{barticle}
\bauthor{\bsnm{{Krieger}}, \binits{A.S.}},
\bauthor{\bsnm{{Timothy}}, \binits{A.F.}},
\bauthor{\bsnm{{Roelof}}, \binits{E.C.}}:
\byear{1973},
\batitle{{A Coronal Hole and Its Identification as the Source of a High
  Velocity Solar Wind Stream}}.
\bjtitle{Sol. Phys.}
\bvolume{29},
\bfpage{505}.
\doiurl{https://doi.org/10.1007/BF00150828}.
\adsurl{1973SoPh...29..505K}.
\end{barticle}
\endbibitem

\bibitem[\protect\citeauthoryear{{Krishna Prasad}, {Banerjee}, and
  {Gupta}}{2011}]{2011A&A...528L...4K}
\begin{barticle}
\bauthor{\bsnm{{Krishna Prasad}}, \binits{S.}},
\bauthor{\bsnm{{Banerjee}}, \binits{D.}},
\bauthor{\bsnm{{Gupta}}, \binits{G.R.}}:
\byear{2011},
\batitle{{Propagating intensity disturbances in polar corona as seen from
  AIA/SDO}}.
\bjtitle{\aap}
\bvolume{528},
\bfpage{L4}.
\doiurl{https://doi.org/10.1051/0004-6361/201016405}.
\adsurl{2011A&A...528L...4K}.
\end{barticle}
\endbibitem

\bibitem[\protect\citeauthoryear{{Kumar} et~al.}{2019}]{2019ApJ...873...93K}
\begin{barticle}
\bauthor{\bsnm{{Kumar}}, \binits{P.}},
\bauthor{\bsnm{{Karpen}}, \binits{J.T.}},
\bauthor{\bsnm{{Antiochos}}, \binits{S.K.}},
\bauthor{\bsnm{{Wyper}}, \binits{P.F.}},
\bauthor{\bsnm{{DeVore}}, \binits{C.R.}},
\bauthor{\bsnm{{DeForest}}, \binits{C.E.}}:
\byear{2019},
\batitle{{Multiwavelength Study of Equatorial Coronal-hole Jets}}.
\bjtitle{\apj}
\bvolume{873},
\bfpage{93}.
\doiurl{https://doi.org/10.3847/1538-4357/ab04af}.
\adsurl{2019ApJ...873...93K}.
\end{barticle}
\endbibitem

\bibitem[\protect\citeauthoryear{{Lamb} et~al.}{2008}]{2008ApJ...674..520L}
\begin{barticle}
\bauthor{\bsnm{{Lamb}}, \binits{D.A.}},
\bauthor{\bsnm{{DeForest}}, \binits{C.E.}},
\bauthor{\bsnm{{Hagenaar}}, \binits{H.J.}},
\bauthor{\bsnm{{Parnell}}, \binits{C.E.}},
\bauthor{\bsnm{{Welsch}}, \binits{B.T.}}:
\byear{2008},
\batitle{{Solar Magnetic Tracking. II. The Apparent Unipolar Origin of
  Quiet-Sun Flux}}.
\bjtitle{\apj}
\bvolume{674},
\bfpage{520}.
\doiurl{https://doi.org/10.1086/524372}.
\adsurl{2008ApJ...674..520L}.
\end{barticle}
\endbibitem

\bibitem[\protect\citeauthoryear{{Lamb} et~al.}{2010}]{2010ApJ...720.1405L}
\begin{barticle}
\bauthor{\bsnm{{Lamb}}, \binits{D.A.}},
\bauthor{\bsnm{{DeForest}}, \binits{C.E.}},
\bauthor{\bsnm{{Hagenaar}}, \binits{H.J.}},
\bauthor{\bsnm{{Parnell}}, \binits{C.E.}},
\bauthor{\bsnm{{Welsch}}, \binits{B.T.}}:
\byear{2010},
\batitle{{Solar Magnetic Tracking. III. Apparent Unipolar Flux Emergence in
  High-resolution Observations}}.
\bjtitle{\apj}
\bvolume{720},
\bfpage{1405}.
\doiurl{https://doi.org/10.1088/0004-637X/720/2/1405}.
\adsurl{2010ApJ...720.1405L}.
\end{barticle}
\endbibitem

\bibitem[\protect\citeauthoryear{{Li} et~al.}{2018}]{2018JGRA..123.7257L}
\begin{barticle}
\bauthor{\bsnm{{Li}}, \binits{X.}},
\bauthor{\bsnm{{Wang}}, \binits{Y.}},
\bauthor{\bsnm{{Liu}}, \binits{R.}},
\bauthor{\bsnm{{Shen}}, \binits{C.}},
\bauthor{\bsnm{{Zhang}}, \binits{Q.}},
\bauthor{\bsnm{{Zhuang}}, \binits{B.}},
\bauthor{\bsnm{{Liu}}, \binits{J.}},
\bauthor{\bsnm{{Chi}}, \binits{Y.}}:
\byear{2018},
\batitle{{Reconstructing Solar Wind Inhomogeneous Structures From Stereoscopic
  Observations in White Light: Small Transients Along the Sun-Earth Line}}.
\bjtitle{Journal of Geophysical Research (Space Physics)}
\bvolume{123},
\bfpage{7257}.
\doiurl{https://doi.org/10.1029/2018JA025485}.
\adsurl{2018JGRA..123.7257L}.
\end{barticle}
\endbibitem

\bibitem[\protect\citeauthoryear{{Li} et~al.}{2020}]{2020JGRA..12527513L}
\begin{barticle}
\bauthor{\bsnm{{Li}}, \binits{X.}},
\bauthor{\bsnm{{Wang}}, \binits{Y.}},
\bauthor{\bsnm{{Liu}}, \binits{R.}},
\bauthor{\bsnm{{Shen}}, \binits{C.}},
\bauthor{\bsnm{{Zhang}}, \binits{Q.}},
\bauthor{\bsnm{{Lyu}}, \binits{S.}},
\bauthor{\bsnm{{Zhuang}}, \binits{B.}},
\bauthor{\bsnm{{Shen}}, \binits{F.}},
\bauthor{\bsnm{{Liu}}, \binits{J.}},
\bauthor{\bsnm{{Chi}}, \binits{Y.}}:
\byear{2020},
\batitle{{Reconstructing Solar Wind Inhomogeneous Structures From Stereoscopic
  Observations in White Light: Solar Wind Transients in 3-D}}.
\bjtitle{Journal of Geophysical Research (Space Physics)}
\bvolume{125},
\bfpage{e27513}.
\doiurl{https://doi.org/10.1029/2019JA027513}.
\adsurl{2020JGRA..12527513L}.
\end{barticle}
\endbibitem

\bibitem[\protect\citeauthoryear{{Lyu}, {Li}, and
  {Wang}}{2020}]{2020AdSpR..66.2251L}
\begin{barticle}
\bauthor{\bsnm{{Lyu}}, \binits{S.}},
\bauthor{\bsnm{{Li}}, \binits{X.}},
\bauthor{\bsnm{{Wang}}, \binits{Y.}}:
\byear{2020},
\batitle{{Optimal stereoscopic angle for reconstructing solar wind
  inhomogeneous structures}}.
\bjtitle{Advances in Space Research}
\bvolume{66},
\bfpage{2251}.
\doiurl{https://doi.org/10.1016/j.asr.2020.07.045}.
\adsurl{2020AdSpR..66.2251L}.
\end{barticle}
\endbibitem

\bibitem[\protect\citeauthoryear{{Madjarska}}{2019}]{2019LRSP...16....2M}
\begin{barticle}
\bauthor{\bsnm{{Madjarska}}, \binits{M.S.}}:
\byear{2019},
\batitle{{Coronal bright points}}.
\bjtitle{Living Reviews in Solar Physics}
\bvolume{16},
\bfpage{2}.
\doiurl{https://doi.org/10.1007/s41116-019-0018-8}.
\adsurl{2019LRSP...16....2M}.
\end{barticle}
\endbibitem

\bibitem[\protect\citeauthoryear{{Madjarska} and
  {Wiegelmann}}{2009}]{2009A&A...503..991M}
\begin{barticle}
\bauthor{\bsnm{{Madjarska}}, \binits{M.S.}},
\bauthor{\bsnm{{Wiegelmann}}, \binits{T.}}:
\byear{2009},
\batitle{{Coronal hole boundaries evolution at small scales. I. EIT 195
  {\r{A}}{\enskip} and TRACE 171 {\r{A}}{\enskip}view}}.
\bjtitle{A\&A}
\bvolume{503},
\bfpage{991}.
\doiurl{https://doi.org/10.1051/0004-6361/200912066}.
\adsurl{2009A&A...503..991M}.
\end{barticle}
\endbibitem

\bibitem[\protect\citeauthoryear{{Madjarska}
  et~al.}{2012}]{2012A&A...545A..67M}
\begin{barticle}
\bauthor{\bsnm{{Madjarska}}, \binits{M.S.}},
\bauthor{\bsnm{{Huang}}, \binits{Z.}},
\bauthor{\bsnm{{Doyle}}, \binits{J.G.}},
\bauthor{\bsnm{{Subramanian}}, \binits{S.}}:
\byear{2012},
\batitle{{Coronal hole boundaries evolution at small scales. III. EIS and SUMER
  views}}.
\bjtitle{A\&A}
\bvolume{545},
\bfpage{A67}.
\doiurl{https://doi.org/10.1051/0004-6361/201219516}.
\adsurl{2012A&A...545A..67M}.
\end{barticle}
\endbibitem

\bibitem[\protect\citeauthoryear{{McGlasson}
  et~al.}{2019}]{2019ApJ...882...16M}
\begin{barticle}
\bauthor{\bsnm{{McGlasson}}, \binits{R.A.}},
\bauthor{\bsnm{{Panesar}}, \binits{N.K.}},
\bauthor{\bsnm{{Sterling}}, \binits{A.C.}},
\bauthor{\bsnm{{Moore}}, \binits{R.L.}}:
\byear{2019},
\batitle{{Magnetic Flux Cancellation as the Trigger Mechanism of Solar Coronal
  Jets}}.
\bjtitle{\apj}
\bvolume{882},
\bfpage{16}.
\doiurl{https://doi.org/10.3847/1538-4357/ab2fe3}.
\adsurl{2019ApJ...882...16M}.
\end{barticle}
\endbibitem

\bibitem[\protect\citeauthoryear{{Mou} et~al.}{2016}]{2016ApJ...818....9M}
\begin{barticle}
\bauthor{\bsnm{{Mou}}, \binits{C.}},
\bauthor{\bsnm{{Huang}}, \binits{Z.}},
\bauthor{\bsnm{{Xia}}, \binits{L.}},
\bauthor{\bsnm{{Madjarska}}, \binits{M.S.}},
\bauthor{\bsnm{{Li}}, \binits{B.}},
\bauthor{\bsnm{{Fu}}, \binits{H.}},
\bauthor{\bsnm{{Jiao}}, \binits{F.}},
\bauthor{\bsnm{{Hou}}, \binits{Z.}}:
\byear{2016},
\batitle{{Magnetic Flux Supplement to Coronal Bright Points}}.
\bjtitle{\apj}
\bvolume{818},
\bfpage{9}.
\doiurl{https://doi.org/10.3847/0004-637X/818/1/9}.
\adsurl{2016ApJ...818....9M}.
\end{barticle}
\endbibitem

\bibitem[\protect\citeauthoryear{{M{\"u}ller}
  et~al.}{2020}]{2020A&A...642A...1M}
\begin{barticle}
\bauthor{\bsnm{{M{\"u}ller}}, \binits{D.}},
\bauthor{\bsnm{{St. Cyr}}, \binits{O.C.}},
\bauthor{\bsnm{{Zouganelis}}, \binits{I.}},
\bauthor{\bsnm{{Gilbert}}, \binits{H.R.}},
\bauthor{\bsnm{{Marsden}}, \binits{R.}},
\bauthor{\bsnm{{Nieves-Chinchilla}}, \binits{T.}},
\bauthor{\bsnm{{Antonucci}}, \binits{E.}},
\bauthor{\bsnm{{Auch{\`e}re}}, \binits{F.}},
\bauthor{\bsnm{{Berghmans}}, \binits{D.}},
\bauthor{\bsnm{{Horbury}}, \binits{T.S.}},
\bauthor{\bsnm{{Howard}}, \binits{R.A.}},
\bauthor{\bsnm{{Krucker}}, \binits{S.}},
\bauthor{\bsnm{{Maksimovic}}, \binits{M.}},
\bauthor{\bsnm{{Owen}}, \binits{C.J.}},
\bauthor{\bsnm{{Rochus}}, \binits{P.}},
\bauthor{\bsnm{{Rodriguez-Pacheco}}, \binits{J.}},
\bauthor{\bsnm{{Romoli}}, \binits{M.}},
\bauthor{\bsnm{{Solanki}}, \binits{S.K.}},
\bauthor{\bsnm{{Bruno}}, \binits{R.}},
\bauthor{\bsnm{{Carlsson}}, \binits{M.}},
\bauthor{\bsnm{{Fludra}}, \binits{A.}},
\bauthor{\bsnm{{Harra}}, \binits{L.}},
\bauthor{\bsnm{{Hassler}}, \binits{D.M.}},
\bauthor{\bsnm{{Livi}}, \binits{S.}},
\bauthor{\bsnm{{Louarn}}, \binits{P.}},
\bauthor{\bsnm{{Peter}}, \binits{H.}},
\bauthor{\bsnm{{Sch{\"u}hle}}, \binits{U.}},
\bauthor{\bsnm{{Teriaca}}, \binits{L.}},
\bauthor{\bsnm{{del Toro Iniesta}}, \binits{J.C.}},
\bauthor{\bsnm{{Wimmer-Schweingruber}}, \binits{R.F.}},
\bauthor{\bsnm{{Marsch}}, \binits{E.}},
\bauthor{\bsnm{{Velli}}, \binits{M.}},
\bauthor{\bsnm{{De Groof}}, \binits{A.}},
\bauthor{\bsnm{{Walsh}}, \binits{A.}},
\bauthor{\bsnm{{Williams}}, \binits{D.}}:
\byear{2020},
\batitle{{The Solar Orbiter mission. Science overview}}.
\bjtitle{\aap}
\bvolume{642},
\bfpage{A1}.
\doiurl{https://doi.org/10.1051/0004-6361/202038467}.
\adsurl{2020A&A...642A...1M}.
\end{barticle}
\endbibitem

\bibitem[\protect\citeauthoryear{{Neugebauer}}{2012}]{2012ApJ...750...50N}
\begin{barticle}
\bauthor{\bsnm{{Neugebauer}}, \binits{M.}}:
\byear{2012},
\batitle{{Evidence for Polar X-Ray Jets as Sources of Microstream Peaks in the
  Solar Wind}}.
\bjtitle{ApJ}
\bvolume{750},
\bfpage{50}.
\doiurl{https://doi.org/10.1088/0004-637X/750/1/50}.
\adsurl{2012ApJ...750...50N}.
\end{barticle}
\endbibitem

\bibitem[\protect\citeauthoryear{{Ofman}, {Nakariakov}, and
  {Sehgal}}{2000}]{2000ApJ...533.1071O}
\begin{barticle}
\bauthor{\bsnm{{Ofman}}, \binits{L.}},
\bauthor{\bsnm{{Nakariakov}}, \binits{V.M.}},
\bauthor{\bsnm{{Sehgal}}, \binits{N.}}:
\byear{2000},
\batitle{{Dissipation of Slow Magnetosonic Waves in Coronal Plumes}}.
\bjtitle{\apj}
\bvolume{533},
\bfpage{1071}.
\doiurl{https://doi.org/10.1086/308691}.
\adsurl{2000ApJ...533.1071O}.
\end{barticle}
\endbibitem

\bibitem[\protect\citeauthoryear{{Panesar}, {Sterling}, and
  {Moore}}{2018}]{2018ApJ...853..189P}
\begin{barticle}
\bauthor{\bsnm{{Panesar}}, \binits{N.K.}},
\bauthor{\bsnm{{Sterling}}, \binits{A.C.}},
\bauthor{\bsnm{{Moore}}, \binits{R.L.}}:
\byear{2018},
\batitle{{Magnetic Flux Cancelation as the Trigger of Solar Coronal Jets in
  Coronal Holes}}.
\bjtitle{\apj}
\bvolume{853},
\bfpage{189}.
\doiurl{https://doi.org/10.3847/1538-4357/aaa3e9}.
\adsurl{2018ApJ...853..189P}.
\end{barticle}
\endbibitem

\bibitem[\protect\citeauthoryear{{Patsourakos} and
  {Vial}}{2000}]{2000A&A...359L...1P}
\begin{barticle}
\bauthor{\bsnm{{Patsourakos}}, \binits{S.}},
\bauthor{\bsnm{{Vial}}, \binits{J.-C.}}:
\byear{2000},
\batitle{{Outflow velocity of interplume regions at the base of Polar Coronal
  Holes}}.
\bjtitle{\aap}
\bvolume{359},
\bfpage{L1}.
\adsurl{2000A&A...359L...1P}.
\end{barticle}
\endbibitem

\bibitem[\protect\citeauthoryear{{Poletto}}{2015}]{2015LRSP...12....7P}
\begin{barticle}
\bauthor{\bsnm{{Poletto}}, \binits{G.}}:
\byear{2015},
\batitle{{Solar Coronal Plumes}}.
\bjtitle{Living Reviews in Solar Physics}
\bvolume{12},
\bfpage{7}.
\doiurl{https://doi.org/10.1007/lrsp-2015-7}.
\adsurl{2015LRSP...12....7P}.
\end{barticle}
\endbibitem

\bibitem[\protect\citeauthoryear{{Popescu}, {Doyle}, and
  {Xia}}{2004}]{2004A&A...421..339P}
\begin{barticle}
\bauthor{\bsnm{{Popescu}}, \binits{M.D.}},
\bauthor{\bsnm{{Doyle}}, \binits{J.G.}},
\bauthor{\bsnm{{Xia}}, \binits{L.D.}}:
\byear{2004},
\batitle{{Network boundary origins of fast solar wind seen in the low
  transition region?}}
\bjtitle{\aap}
\bvolume{421},
\bfpage{339}.
\doiurl{https://doi.org/10.1051/0004-6361:20034348}.
\adsurl{2004A&A...421..339P}.
\end{barticle}
\endbibitem

\bibitem[\protect\citeauthoryear{{Pucci} et~al.}{2014}]{2014ApJ...793...86P}
\begin{barticle}
\bauthor{\bsnm{{Pucci}}, \binits{S.}},
\bauthor{\bsnm{{Poletto}}, \binits{G.}},
\bauthor{\bsnm{{Sterling}}, \binits{A.C.}},
\bauthor{\bsnm{{Romoli}}, \binits{M.}}:
\byear{2014},
\batitle{{Birth, Life, and Death of a Solar Coronal Plume}}.
\bjtitle{\apj}
\bvolume{793},
\bfpage{86}.
\doiurl{https://doi.org/10.1088/0004-637X/793/2/86}.
\adsurl{2014ApJ...793...86P}.
\end{barticle}
\endbibitem

\bibitem[\protect\citeauthoryear{{Qi} et~al.}{2019}]{2019SoPh..294...92Q}
\begin{barticle}
\bauthor{\bsnm{{Qi}}, \binits{Y.}},
\bauthor{\bsnm{{Huang}}, \binits{Z.}},
\bauthor{\bsnm{{Xia}}, \binits{L.}},
\bauthor{\bsnm{{Li}}, \binits{B.}},
\bauthor{\bsnm{{Fu}}, \binits{H.}},
\bauthor{\bsnm{{Liu}}, \binits{W.}},
\bauthor{\bsnm{{Sun}}, \binits{M.}},
\bauthor{\bsnm{{Hou}}, \binits{Z.}}:
\byear{2019},
\batitle{{On the Relation Between Transition Region Network Jets and Coronal
  Plumes}}.
\bjtitle{\solphys}
\bvolume{294},
\bfpage{92}.
\doiurl{https://doi.org/10.1007/s11207-019-1484-9}.
\adsurl{2019SoPh..294...92Q}.
\end{barticle}
\endbibitem

\bibitem[\protect\citeauthoryear{{Raouafi} and
  {Stenborg}}{2014}]{2014ApJ...787..118R}
\begin{barticle}
\bauthor{\bsnm{{Raouafi}}, \binits{N.-E.}},
\bauthor{\bsnm{{Stenborg}}, \binits{G.}}:
\byear{2014},
\batitle{{Role of Transients in the Sustainability of Solar Coronal Plumes}}.
\bjtitle{\apj}
\bvolume{787},
\bfpage{118}.
\doiurl{https://doi.org/10.1088/0004-637X/787/2/118}.
\adsurl{2014ApJ...787..118R}.
\end{barticle}
\endbibitem

\bibitem[\protect\citeauthoryear{{Rieutord} and
  {Rincon}}{2010}]{2010LRSP....7....2R}
\begin{barticle}
\bauthor{\bsnm{{Rieutord}}, \binits{M.}},
\bauthor{\bsnm{{Rincon}}, \binits{F.}}:
\byear{2010},
\batitle{{The Sun's Supergranulation}}.
\bjtitle{Living Reviews in Solar Physics}
\bvolume{7},
\bfpage{2}.
\doiurl{https://doi.org/10.12942/lrsp-2010-2}.
\adsurl{2010LRSP....7....2R}.
\end{barticle}
\endbibitem

\bibitem[\protect\citeauthoryear{{Shen} et~al.}{2011}]{2011ApJ...735L..43S}
\begin{barticle}
\bauthor{\bsnm{{Shen}}, \binits{Y.}},
\bauthor{\bsnm{{Liu}}, \binits{Y.}},
\bauthor{\bsnm{{Su}}, \binits{J.}},
\bauthor{\bsnm{{Ibrahim}}, \binits{A.}}:
\byear{2011},
\batitle{{Kinematics and Fine Structure of an Unwinding Polar Jet Observed by
  the Solar Dynamic Observatory/Atmospheric Imaging Assembly}}.
\bjtitle{\apjl}
\bvolume{735},
\bfpage{L43}.
\doiurl{https://doi.org/10.1088/2041-8205/735/2/L43}.
\adsurl{2011ApJ...735L..43S}.
\end{barticle}
\endbibitem

\bibitem[\protect\citeauthoryear{{Spice Consortium}
  et~al.}{2020}]{2020A&A...642A..14S}
\begin{barticle}
\bauthor{\bsnm{{Spice Consortium}}},
\bauthor{\bsnm{{Anderson}}, \binits{M.}},
\bauthor{\bsnm{{Appourchaux}}, \binits{T.}},
\bauthor{\bsnm{{Auch{\`e}re}}, \binits{F.}},
\bauthor{\bsnm{{Aznar Cuadrado}}, \binits{R.}},
\bauthor{\bsnm{{Barbay}}, \binits{J.}},
\bauthor{\bsnm{{Baudin}}, \binits{F.}},
\bauthor{\bsnm{{Beardsley}}, \binits{S.}},
\bauthor{\bsnm{{Bocchialini}}, \binits{K.}},
\bauthor{\bsnm{{Borgo}}, \binits{B.}},
\bauthor{\bsnm{{Bruzzi}}, \binits{D.}},
\bauthor{\bsnm{{Buchlin}}, \binits{E.}},
\bauthor{\bsnm{{Burton}}, \binits{G.}},
\bauthor{\bsnm{{B{\"u}chel}}, \binits{V.}},
\bauthor{\bsnm{{Caldwell}}, \binits{M.}},
\bauthor{\bsnm{{Caminade}}, \binits{S.}},
\bauthor{\bsnm{{Carlsson}}, \binits{M.}},
\bauthor{\bsnm{{Curdt}}, \binits{W.}},
\bauthor{\bsnm{{Davenne}}, \binits{J.}},
\bauthor{\bsnm{{Davila}}, \binits{J.}},
\bauthor{\bsnm{{Deforest}}, \binits{C.E.}},
\bauthor{\bsnm{{Del Zanna}}, \binits{G.}},
\bauthor{\bsnm{{Drummond}}, \binits{D.}},
\bauthor{\bsnm{{Dubau}}, \binits{J.}},
\bauthor{\bsnm{{Dumesnil}}, \binits{C.}},
\bauthor{\bsnm{{Dunn}}, \binits{G.}},
\bauthor{\bsnm{{Eccleston}}, \binits{P.}},
\bauthor{\bsnm{{Fludra}}, \binits{A.}},
\bauthor{\bsnm{{Fredvik}}, \binits{T.}},
\bauthor{\bsnm{{Gabriel}}, \binits{A.}},
\bauthor{\bsnm{{Giunta}}, \binits{A.}},
\bauthor{\bsnm{{Gottwald}}, \binits{A.}},
\bauthor{\bsnm{{Griffin}}, \binits{D.}},
\bauthor{\bsnm{{Grundy}}, \binits{T.}},
\bauthor{\bsnm{{Guest}}, \binits{S.}},
\bauthor{\bsnm{{Gyo}}, \binits{M.}},
\bauthor{\bsnm{{Haberreiter}}, \binits{M.}},
\bauthor{\bsnm{{Hansteen}}, \binits{V.}},
\bauthor{\bsnm{{Harrison}}, \binits{R.}},
\bauthor{\bsnm{{Hassler}}, \binits{D.M.}},
\bauthor{\bsnm{{Haugan}}, \binits{S.V.H.}},
\bauthor{\bsnm{{Howe}}, \binits{C.}},
\bauthor{\bsnm{{Janvier}}, \binits{M.}},
\bauthor{\bsnm{{Klein}}, \binits{R.}},
\bauthor{\bsnm{{Koller}}, \binits{S.}},
\bauthor{\bsnm{{Kucera}}, \binits{T.A.}},
\bauthor{\bsnm{{Kouliche}}, \binits{D.}},
\bauthor{\bsnm{{Marsch}}, \binits{E.}},
\bauthor{\bsnm{{Marshall}}, \binits{A.}},
\bauthor{\bsnm{{Marshall}}, \binits{G.}},
\bauthor{\bsnm{{Matthews}}, \binits{S.A.}},
\bauthor{\bsnm{{McQuirk}}, \binits{C.}},
\bauthor{\bsnm{{Meining}}, \binits{S.}},
\bauthor{\bsnm{{Mercier}}, \binits{C.}},
\bauthor{\bsnm{{Morris}}, \binits{N.}},
\bauthor{\bsnm{{Morse}}, \binits{T.}},
\bauthor{\bsnm{{Munro}}, \binits{G.}},
\bauthor{\bsnm{{Parenti}}, \binits{S.}},
\bauthor{\bsnm{{Pastor-Santos}}, \binits{C.}},
\bauthor{\bsnm{{Peter}}, \binits{H.}},
\bauthor{\bsnm{{Pfiffner}}, \binits{D.}},
\bauthor{\bsnm{{Phelan}}, \binits{P.}},
\bauthor{\bsnm{{Philippon}}, \binits{A.}},
\bauthor{\bsnm{{Richards}}, \binits{A.}},
\bauthor{\bsnm{{Rogers}}, \binits{K.}},
\bauthor{\bsnm{{Sawyer}}, \binits{C.}},
\bauthor{\bsnm{{Schlatter}}, \binits{P.}},
\bauthor{\bsnm{{Schmutz}}, \binits{W.}},
\bauthor{\bsnm{{Sch{\"u}hle}}, \binits{U.}},
\bauthor{\bsnm{{Shaughnessy}}, \binits{B.}},
\bauthor{\bsnm{{Sidher}}, \binits{S.}},
\bauthor{\bsnm{{Solanki}}, \binits{S.K.}},
\bauthor{\bsnm{{Speight}}, \binits{R.}},
\bauthor{\bsnm{{Spescha}}, \binits{M.}},
\bauthor{\bsnm{{Szwec}}, \binits{N.}},
\bauthor{\bsnm{{Tamiatto}}, \binits{C.}},
\bauthor{\bsnm{{Teriaca}}, \binits{L.}},
\bauthor{\bsnm{{Thompson}}, \binits{W.}},
\bauthor{\bsnm{{Tosh}}, \binits{I.}},
\bauthor{\bsnm{{Tustain}}, \binits{S.}},
\bauthor{\bsnm{{Vial}}, \binits{J.-C.}},
\bauthor{\bsnm{{Walls}}, \binits{B.}},
\bauthor{\bsnm{{Waltham}}, \binits{N.}},
\bauthor{\bsnm{{Wimmer-Schweingruber}}, \binits{R.}},
\bauthor{\bsnm{{Woodward}}, \binits{S.}},
\bauthor{\bsnm{{Young}}, \binits{P.}},
\bauthor{\bsnm{{de Groof}}, \binits{A.}},
\bauthor{\bsnm{{Pacros}}, \binits{A.}},
\bauthor{\bsnm{{Williams}}, \binits{D.}},
\bauthor{\bsnm{{M{\"u}ller}}, \binits{D.}}:
\byear{2020},
\batitle{{The Solar Orbiter SPICE instrument. An extreme UV imaging
  spectrometer}}.
\bjtitle{\aap}
\bvolume{642},
\bfpage{A14}.
\doiurl{https://doi.org/10.1051/0004-6361/201935574}.
\adsurl{2020A&A...642A..14S}.
\end{barticle}
\endbibitem

\bibitem[\protect\citeauthoryear{{Sterling}}{2000}]{2000SoPh..196...79S}
\begin{barticle}
\bauthor{\bsnm{{Sterling}}, \binits{A.C.}}:
\byear{2000},
\batitle{{Solar Spicules: A Review of Recent Models and Targets for Future
  Observations - (Invited Review)}}.
\bjtitle{\solphys}
\bvolume{196},
\bfpage{79}.
\doiurl{https://doi.org/10.1023/A:1005213923962}.
\adsurl{2000SoPh..196...79S}.
\end{barticle}
\endbibitem

\bibitem[\protect\citeauthoryear{{Subramanian}, {Madjarska}, and
  {Doyle}}{2010}]{2010A&A...516A..50S}
\begin{barticle}
\bauthor{\bsnm{{Subramanian}}, \binits{S.}},
\bauthor{\bsnm{{Madjarska}}, \binits{M.S.}},
\bauthor{\bsnm{{Doyle}}, \binits{J.G.}}:
\byear{2010},
\batitle{{Coronal hole boundaries evolution at small scales. II. XRT view. Can
  small-scale outflows at CHBs be a source of the slow solar wind}}.
\bjtitle{A\&A}
\bvolume{516},
\bfpage{A50}.
\doiurl{https://doi.org/10.1051/0004-6361/200913624}.
\adsurl{2010A&A...516A..50S}.
\end{barticle}
\endbibitem

\bibitem[\protect\citeauthoryear{{Teriaca} et~al.}{2003}]{2003ApJ...588..566T}
\begin{barticle}
\bauthor{\bsnm{{Teriaca}}, \binits{L.}},
\bauthor{\bsnm{{Poletto}}, \binits{G.}},
\bauthor{\bsnm{{Romoli}}, \binits{M.}},
\bauthor{\bsnm{{Biesecker}}, \binits{D.A.}}:
\byear{2003},
\batitle{{The Nascent Solar Wind: Origin and Acceleration}}.
\bjtitle{\apj}
\bvolume{588},
\bfpage{566}.
\doiurl{https://doi.org/10.1086/368409}.
\adsurl{2003ApJ...588..566T}.
\end{barticle}
\endbibitem

\bibitem[\protect\citeauthoryear{{Thompson}}{2006}]{2006A&A...449..791T}
\begin{barticle}
\bauthor{\bsnm{{Thompson}}, \binits{W.T.}}:
\byear{2006},
\batitle{{Coordinate systems for solar image data}}.
\bjtitle{\aap}
\bvolume{449},
\bfpage{791}.
\doiurl{https://doi.org/10.1051/0004-6361:20054262}.
\adsurl{2006A&A...449..791T}.
\end{barticle}
\endbibitem

\bibitem[\protect\citeauthoryear{{Thompson} and
  {Wei}}{2010}]{2010SoPh..261..215T}
\begin{barticle}
\bauthor{\bsnm{{Thompson}}, \binits{W.T.}},
\bauthor{\bsnm{{Wei}}, \binits{K.}}:
\byear{2010},
\batitle{{Use of the FITS World Coordinate System by STEREO/SECCHI}}.
\bjtitle{\solphys}
\bvolume{261},
\bfpage{215}.
\doiurl{https://doi.org/10.1007/s11207-009-9476-9}.
\adsurl{2010SoPh..261..215T}.
\end{barticle}
\endbibitem

\bibitem[\protect\citeauthoryear{{Tian} et~al.}{2011a}]{2011ApJ...736..130T}
\begin{barticle}
\bauthor{\bsnm{{Tian}}, \binits{H.}},
\bauthor{\bsnm{{McIntosh}}, \binits{S.W.}},
\bauthor{\bsnm{{Habbal}}, \binits{S.R.}},
\bauthor{\bsnm{{He}}, \binits{J.}}:
\byear{2011}a,
\batitle{{Observation of High-speed Outflow on Plume-like Structures of the
  Quiet Sun and Coronal Holes with Solar Dynamics Observatory/Atmospheric
  Imaging Assembly}}.
\bjtitle{\apj}
\bvolume{736},
\bfpage{130}.
\doiurl{https://doi.org/10.1088/0004-637X/736/2/130}.
\adsurl{2011ApJ...736..130T}.
\end{barticle}
\endbibitem

\bibitem[\protect\citeauthoryear{{Tian} et~al.}{2011b}]{2011ApJ...738...18T}
\begin{barticle}
\bauthor{\bsnm{{Tian}}, \binits{H.}},
\bauthor{\bsnm{{McIntosh}}, \binits{S.W.}},
\bauthor{\bsnm{{De Pontieu}}, \binits{B.}},
\bauthor{\bsnm{{Mart{\'\i}nez-Sykora}}, \binits{J.}},
\bauthor{\bsnm{{Sechler}}, \binits{M.}},
\bauthor{\bsnm{{Wang}}, \binits{X.}}:
\byear{2011}b,
\batitle{{Two Components of the Solar Coronal Emission Revealed by
  Extreme-ultraviolet Spectroscopic Observations}}.
\bjtitle{\apj}
\bvolume{738},
\bfpage{18}.
\doiurl{https://doi.org/10.1088/0004-637X/738/1/18}.
\adsurl{2011ApJ...738...18T}.
\end{barticle}
\endbibitem

\bibitem[\protect\citeauthoryear{{Tian} et~al.}{2012}]{2012ApJ...759..144T}
\begin{barticle}
\bauthor{\bsnm{{Tian}}, \binits{H.}},
\bauthor{\bsnm{{McIntosh}}, \binits{S.W.}},
\bauthor{\bsnm{{Wang}}, \binits{T.}},
\bauthor{\bsnm{{Ofman}}, \binits{L.}},
\bauthor{\bsnm{{De Pontieu}}, \binits{B.}},
\bauthor{\bsnm{{Innes}}, \binits{D.E.}},
\bauthor{\bsnm{{Peter}}, \binits{H.}}:
\byear{2012},
\batitle{{Persistent Doppler Shift Oscillations Observed with Hinode/EIS in the
  Solar Corona: Spectroscopic Signatures of Alfv{\'e}nic Waves and Recurring
  Upflows}}.
\bjtitle{\apj}
\bvolume{759},
\bfpage{144}.
\doiurl{https://doi.org/10.1088/0004-637X/759/2/144}.
\adsurl{2012ApJ...759..144T}.
\end{barticle}
\endbibitem

\bibitem[\protect\citeauthoryear{{Tu} et~al.}{2005}]{2005Sci...308..519T}
\begin{barticle}
\bauthor{\bsnm{{Tu}}, \binits{C.-Y.}},
\bauthor{\bsnm{{Zhou}}, \binits{C.}},
\bauthor{\bsnm{{Marsch}}, \binits{E.}},
\bauthor{\bsnm{{Xia}}, \binits{L.-D.}},
\bauthor{\bsnm{{Zhao}}, \binits{L.}},
\bauthor{\bsnm{{Wang}}, \binits{J.-X.}},
\bauthor{\bsnm{{Wilhelm}}, \binits{K.}}:
\byear{2005},
\batitle{{Solar Wind Origin in Coronal Funnels}}.
\bjtitle{Science}
\bvolume{308},
\bfpage{519}.
\doiurl{https://doi.org/10.1126/science.1109447}.
\adsurl{2005Sci...308..519T}.
\end{barticle}
\endbibitem

\bibitem[\protect\citeauthoryear{{Wang}}{1998}]{1998ApJ...501L.145W}
\begin{barticle}
\bauthor{\bsnm{{Wang}}, \binits{Y.-M.}}:
\byear{1998},
\batitle{{Network Activity and the Evaporative Formation of Polar Plumes}}.
\bjtitle{\apjl}
\bvolume{501},
\bfpage{L145}.
\doiurl{https://doi.org/10.1086/311445}.
\adsurl{1998ApJ...501L.145W}.
\end{barticle}
\endbibitem

\bibitem[\protect\citeauthoryear{{Wang} and
  {Muglach}}{2008}]{2008SoPh..249...17W}
\begin{barticle}
\bauthor{\bsnm{{Wang}}, \binits{Y.-M.}},
\bauthor{\bsnm{{Muglach}}, \binits{K.}}:
\byear{2008},
\batitle{{Observations of Low-Latitude Coronal Plumes}}.
\bjtitle{\solphys}
\bvolume{249},
\bfpage{17}.
\doiurl{https://doi.org/10.1007/s11207-008-9171-2}.
\adsurl{2008SoPh..249...17W}.
\end{barticle}
\endbibitem

\bibitem[\protect\citeauthoryear{{Wang}, {Warren}, and
  {Muglach}}{2016}]{2016ApJ...818..203W}
\begin{barticle}
\bauthor{\bsnm{{Wang}}, \binits{Y.-M.}},
\bauthor{\bsnm{{Warren}}, \binits{H.P.}},
\bauthor{\bsnm{{Muglach}}, \binits{K.}}:
\byear{2016},
\batitle{{Converging Supergranular Flows and the Formation of Coronal Plumes}}.
\bjtitle{\apj}
\bvolume{818},
\bfpage{203}.
\doiurl{https://doi.org/10.3847/0004-637X/818/2/203}.
\adsurl{2016ApJ...818..203W}.
\end{barticle}
\endbibitem

\bibitem[\protect\citeauthoryear{{Wang} et~al.}{2020a}]{2020ScChE..63.1699W}
\begin{barticle}
\bauthor{\bsnm{{Wang}}, \binits{Y.}},
\bauthor{\bsnm{{Ji}}, \binits{H.}},
\bauthor{\bsnm{{Wang}}, \binits{Y.}},
\bauthor{\bsnm{{Xia}}, \binits{L.}},
\bauthor{\bsnm{{Shen}}, \binits{C.}},
\bauthor{\bsnm{{Guo}}, \binits{J.}},
\bauthor{\bsnm{{Zhang}}, \binits{Q.}},
\bauthor{\bsnm{{Huang}}, \binits{Z.}},
\bauthor{\bsnm{{Liu}}, \binits{K.}},
\bauthor{\bsnm{{Li}}, \binits{X.}},
\bauthor{\bsnm{{Liu}}, \binits{R.}},
\bauthor{\bsnm{{Wang}}, \binits{J.}},
\bauthor{\bsnm{{Wang}}, \binits{S.}}:
\byear{2020}a,
\batitle{{Concept of the solar ring mission: An overview}}.
\bjtitle{Science in China E: Technological Sciences}
\bvolume{63},
\bfpage{1699}.
\doiurl{https://doi.org/10.1007/s11431-020-1603-2}.
\adsurl{2020ScChE..63.1699W}.
\end{barticle}
\endbibitem

\bibitem[\protect\citeauthoryear{{Wang} et~al.}{2020b}]{wang2020scic}
\begin{botherref}
\oauthor{\bsnm{{Wang}}, \binits{Y.}},
\oauthor{\bsnm{{Chen}}, \binits{X.}},
\oauthor{\bsnm{{Wang}}, \binits{P.}},
\oauthor{\bsnm{{Qiu}}, \binits{C.}},
\oauthor{\bsnm{{Wang}}, \binits{Y.}},
\oauthor{\bsnm{{Zhang}}, \binits{Y.}}:
2020b,
{Concept of the Solar Ring mission: Preliminary design and mission profile}.
\textit{Sci. China Technol. Sci.}
\doiurl{https://doi.org/10.1007/s11431-020-1612-y}.
\end{botherref}
\endbibitem

\bibitem[\protect\citeauthoryear{{Wilhelm} et~al.}{2000}]{2000A&A...353..749W}
\begin{barticle}
\bauthor{\bsnm{{Wilhelm}}, \binits{K.}},
\bauthor{\bsnm{{Dammasch}}, \binits{I.E.}},
\bauthor{\bsnm{{Marsch}}, \binits{E.}},
\bauthor{\bsnm{{Hassler}}, \binits{D.M.}}:
\byear{2000},
\batitle{{On the source regions of the fast solar wind in polar coronal
  holes}}.
\bjtitle{\aap}
\bvolume{353},
\bfpage{749}.
\adsurl{2000A&A...353..749W}.
\end{barticle}
\endbibitem

\bibitem[\protect\citeauthoryear{{Wilhelm} et~al.}{2011}]{2011A&ARv..19...35W}
\begin{barticle}
\bauthor{\bsnm{{Wilhelm}}, \binits{K.}},
\bauthor{\bsnm{{Abbo}}, \binits{L.}},
\bauthor{\bsnm{{Auch{\`e}re}}, \binits{F.}},
\bauthor{\bsnm{{Barbey}}, \binits{N.}},
\bauthor{\bsnm{{Feng}}, \binits{L.}},
\bauthor{\bsnm{{Gabriel}}, \binits{A.H.}},
\bauthor{\bsnm{{Giordano}}, \binits{S.}},
\bauthor{\bsnm{{Imada}}, \binits{S.}},
\bauthor{\bsnm{{Llebaria}}, \binits{A.}},
\bauthor{\bsnm{{Matthaeus}}, \binits{W.H.}},
\bauthor{\bsnm{{Poletto}}, \binits{G.}},
\bauthor{\bsnm{{Raouafi}}, \binits{N.-E.}},
\bauthor{\bsnm{{Suess}}, \binits{S.T.}},
\bauthor{\bsnm{{Teriaca}}, \binits{L.}},
\bauthor{\bsnm{{Wang}}, \binits{Y.-M.}}:
\byear{2011},
\batitle{{Morphology, dynamics and plasma parameters of plumes and inter-plume
  regions in solar coronal holes}}.
\bjtitle{\aapr}
\bvolume{19},
\bfpage{35}.
\doiurl{https://doi.org/10.1007/s00159-011-0035-7}.
\adsurl{2011A&ARv..19...35W}.
\end{barticle}
\endbibitem

\bibitem[\protect\citeauthoryear{{Wuelser} et~al.}{2004}]{2004SPIE.5171..111W}
\begin{bchapter}
\bauthor{\bsnm{{Wuelser}}, \binits{J.-P.}},
\bauthor{\bsnm{{Lemen}}, \binits{J.R.}},
\bauthor{\bsnm{{Tarbell}}, \binits{T.D.}},
\bauthor{\bsnm{{Wolfson}}, \binits{C.J.}},
\bauthor{\bsnm{{Cannon}}, \binits{J.C.}},
\bauthor{\bsnm{{Carpenter}}, \binits{B.A.}},
\bauthor{\bsnm{{Duncan}}, \binits{D.W.}},
\bauthor{\bsnm{{Gradwohl}}, \binits{G.S.}},
\bauthor{\bsnm{{Meyer}}, \binits{S.B.}},
\bauthor{\bsnm{{Moore}}, \binits{A.S.}},
\bauthor{\bsnm{{Navarro}}, \binits{R.L.}},
\bauthor{\bsnm{{Pearson}}, \binits{J.D.}},
\bauthor{\bsnm{{Rossi}}, \binits{G.R.}},
\bauthor{\bsnm{{Springer}}, \binits{L.A.}},
\bauthor{\bsnm{{Howard}}, \binits{R.A.}},
\bauthor{\bsnm{{Moses}}, \binits{J.D.}},
\bauthor{\bsnm{{Newmark}}, \binits{J.S.}},
\bauthor{\bsnm{{Delaboudiniere}}, \binits{J.-P.}},
\bauthor{\bsnm{{Artzner}}, \binits{G.E.}},
\bauthor{\bsnm{{Auchere}}, \binits{F.}},
\bauthor{\bsnm{{Bougnet}}, \binits{M.}},
\bauthor{\bsnm{{Bouyries}}, \binits{P.}},
\bauthor{\bsnm{{Bridou}}, \binits{F.}},
\bauthor{\bsnm{{Clotaire}}, \binits{J.-Y.}},
\bauthor{\bsnm{{Colas}}, \binits{G.}},
\bauthor{\bsnm{{Delmotte}}, \binits{F.}},
\bauthor{\bsnm{{Jerome}}, \binits{A.}},
\bauthor{\bsnm{{Lamare}}, \binits{M.}},
\bauthor{\bsnm{{Mercier}}, \binits{R.}},
\bauthor{\bsnm{{Mullot}}, \binits{M.}},
\bauthor{\bsnm{{Ravet}}, \binits{M.-F.}},
\bauthor{\bsnm{{Song}}, \binits{X.}},
\bauthor{\bsnm{{Bothmer}}, \binits{V.}},
\bauthor{\bsnm{{Deutsch}}, \binits{W.}}:
\byear{2004},
\bctitle{{EUVI: the STEREO-SECCHI extreme ultraviolet imager}}.
In: \beditor{\bsnm{{Fineschi}}, \binits{S.}},
\beditor{\bsnm{{Gummin}}, \binits{M.A.}} (eds.)
\bbtitle{Telescopes and Instrumentation for Solar Astrophysics},
\bsertitle{Society of Photo-Optical Instrumentation Engineers (SPIE) Conference
  Series}
\bseriesno{5171},
\bfpage{111}.
\doiurl{https://doi.org/10.1117/12.506877}.
\adsurl{2004SPIE.5171..111W}.
\end{bchapter}
\endbibitem

\bibitem[\protect\citeauthoryear{{Xia}, {Marsch}, and
  {Curdt}}{2003}]{2003A&A...399L...5X}
\begin{barticle}
\bauthor{\bsnm{{Xia}}, \binits{L.D.}},
\bauthor{\bsnm{{Marsch}}, \binits{E.}},
\bauthor{\bsnm{{Curdt}}, \binits{W.}}:
\byear{2003},
\batitle{{On the outflow in an equatorial coronal hole}}.
\bjtitle{\aap}
\bvolume{399},
\bfpage{L5}.
\doiurl{https://doi.org/10.1051/0004-6361:20030016}.
\adsurl{2003A&A...399L...5X}.
\end{barticle}
\endbibitem

\bibitem[\protect\citeauthoryear{{Xia}, {Marsch}, and
  {Wilhelm}}{2004}]{2004A&A...424.1025X}
\begin{barticle}
\bauthor{\bsnm{{Xia}}, \binits{L.D.}},
\bauthor{\bsnm{{Marsch}}, \binits{E.}},
\bauthor{\bsnm{{Wilhelm}}, \binits{K.}}:
\byear{2004},
\batitle{{On the network structures in solar equatorial coronal holes.
  Observations of SUMER and MDI on SOHO}}.
\bjtitle{\aap}
\bvolume{424},
\bfpage{1025}.
\doiurl{https://doi.org/10.1051/0004-6361:20047027}.
\adsurl{2004A&A...424.1025X}.
\end{barticle}
\endbibitem

\bibitem[\protect\citeauthoryear{{Xie} et~al.}{2009}]{2009A&A...505..801X}
\begin{barticle}
\bauthor{\bsnm{{Xie}}, \binits{Z.X.}},
\bauthor{\bsnm{{Yu}}, \binits{D.R.}},
\bauthor{\bsnm{{Zhang}}, \binits{J.}},
\bauthor{\bsnm{{Yang}}, \binits{S.H.}},
\bauthor{\bsnm{{Hu}}, \binits{Q.H.}}:
\byear{2009},
\batitle{{Properties of magnetic elements in the quiet Sun using the
  marker-controlled watershed method}}.
\bjtitle{\aap}
\bvolume{505},
\bfpage{801}.
\doiurl{https://doi.org/10.1051/0004-6361/200810946}.
\adsurl{2009A&A...505..801X}.
\end{barticle}
\endbibitem

\bibitem[\protect\citeauthoryear{{Yang} et~al.}{2011}]{2011SCPMA..54.1906Y}
\begin{barticle}
\bauthor{\bsnm{{Yang}}, \binits{S.}},
\bauthor{\bsnm{{Zhang}}, \binits{J.}},
\bauthor{\bsnm{{Zhang}}, \binits{Z.}},
\bauthor{\bsnm{{Zhao}}, \binits{Z.}},
\bauthor{\bsnm{{Liu}}, \binits{Y.}},
\bauthor{\bsnm{{Song}}, \binits{Q.}},
\bauthor{\bsnm{{Yang}}, \binits{S.}},
\bauthor{\bsnm{{Bao}}, \binits{X.}},
\bauthor{\bsnm{{Li}}, \binits{L.}},
\bauthor{\bsnm{{Chu}}, \binits{Z.}},
\bauthor{\bsnm{{Li}}, \binits{T.}}:
\byear{2011},
\batitle{{Polar plumes observed at the total solar eclipse in 2009}}.
\bjtitle{Science China Physics, Mechanics, and Astronomy}
\bvolume{54},
\bfpage{1906}.
\doiurl{https://doi.org/10.1007/s11433-011-4471-1}.
\adsurl{2011SCPMA..54.1906Y}.
\end{barticle}
\endbibitem

\bibitem[\protect\citeauthoryear{{Zangrilli} and
  {Giordano}}{2020}]{2020A&A...643A.104Z}
\begin{barticle}
\bauthor{\bsnm{{Zangrilli}}, \binits{L.}},
\bauthor{\bsnm{{Giordano}}, \binits{S.M.}}:
\byear{2020},
\batitle{{Contribution of polar plumes to fast solar wind}}.
\bjtitle{\aap}
\bvolume{643},
\bfpage{A104}.
\doiurl{https://doi.org/10.1051/0004-6361/202037653}.
\adsurl{2020A&A...643A.104Z}.
\end{barticle}
\endbibitem

\end{thebibliography}

\end{article}
\end{document}